\documentclass{elsart}

\usepackage{psfig,epsfig,latexsym,amssymb}

\def\Lattice{
\begin{picture}(130.,40.)
\put(0,0){\line(1,0){125.}}
\put(0,25){\line(1,0){125.}}  
\put(130.,0.){\makebox(0.,0.){$S$}}
\put(130.,25.){\makebox(0.,0.){$T$}}
\put(-0.5,0){\line(0,1){25.}}
\put(0.5,0){\line(0,1){25.}}
\put(0,0){\line(0,1){25.}}
\put(125,0){\line(0,1){25.}}
\put(125.5,0){\line(0,1){25.}}
\put(124.5,0){\line(0,1){25.}}
\put(25,0){\line(0,1){25.}}  
\put(50,0){\line(0,1){25.}}
\put(75,0){\line(0,1){25.}}  
\put(100,0){\line(0,1){25.}}
\put(0.,28.0){\makebox(0.,0.){$1$}}
\put(25.,28.0){\makebox(0.,0.){$2$}}
\put(125.,28.0){\makebox(0.,0.){$L$}}
\put(0.,0.){\makebox(0.,0.){$\bullet$}}
\put(0.,25.){\makebox(0.,0.){$\bullet$}}
\put(25.,0.){\makebox(0.,0.){$\bullet$}}
\put(25.,25.){\makebox(0.,0.){$\bullet$}}
\put(50.,0.){\makebox(0.,0.){$\bullet$}}
\put(50.,25.){\makebox(0.,0.){$\bullet$}}
\put(75.,0.){\makebox(0.,0.){$\bullet$}}
\put(75.,25.){\makebox(0.,0.){$\bullet$}}
\put(100.,0.){\makebox(0.,0.){$\bullet$}}
\put(100.,25.){\makebox(0.,0.){$\bullet$}}
\put(125.,0.){\makebox(0.,0.){$\bullet$}}
\put(125.,25.){\makebox(0.,0.){$\bullet$}}
\put(53.5,12.0){\makebox(0.,0.){$J_{\perp}$}}
\put(60.5,21.0){\makebox(0.,0.){$J_{\parallel}$}}
\put(60.5,3.0){\makebox(0.,0.){$J_{\parallel}$}}
\put(4.,11.5){\makebox(0.,0.){$U_-$}}
\put(119.5,11.5){\makebox(0.,0.){$U_+$}}
\end{picture}}

\begin{document}

\begin{frontmatter}

  \title{Thermodynamic properties of
  an integrable quantum spin ladder with boundary impurities}
\author[ANU1,ANU2]{M.T. Batchelor},
\author[ANU1,ANU2]{X.-W.~Guan},
\author[PortoAlegre]{A.~Foerster},
\author[PortoAlegre]{ A.P. Tonel},
\author[Queensland]{H.-Q.~Zhou}
\address[ANU1]{Department of Theoretical Physics, Research School of Physical
  Sciences and Engineering,
 Australian National University, Canberra ACT 0200,  Australia}
\address[ANU2]{Centre for Mathematics and its applications, School of Mathematical Sciences,\\
Australian National University, Canberra ACT 0200,  Australia}
\address[PortoAlegre]{Instituto de Fisica
  da UFRGS, Av.\ Bento Goncalves, 9500,\\ Porto Alegre, 91501-970,
  Brazil} 
\address[Queensland]{Centre for Mathematical Physics, 
School of Physical Sciences, \\ The University of Queensland, 4072, Australia}

\begin{abstract}
 An integrable quantum spin ladder based on the SU(4) symmetry 
algebra with boundary defects is studied in the framework of 
boundary integrability. Five nontrivial solutions of the 
reflection equations lead to different boundary impurities. 
In each case the energy spectrum is determined using the 
quantum inverse scattering method. The thermodynamic properties 
are investigated by means of the thermodynamic Bethe ansatz.
In particular, the susceptibility and the magnetization of the
model in the vicinity of the critical points are derived along
with differing magnetic properites for antiferromagnetic and
ferromagnetic impurity couplings at the edges. The results are
applicable to the strong coupling ladder compounds, such as 
Cu$_{2}$(C$_5$H$_{12}$N$_2$)$_2$Cl$_4$.

\end{abstract}
\begin{keyword}
  Yang-Baxter equation; reflection equations; thermodynamic 
  Bethe ansatz;  impurity effects, quantum spin ladders
\PACS{75.10.Jm; 75.30.Kz; 75.40.Cx}
\end{keyword}

\end{frontmatter}

\section{Introduction}
\label{sec1}

Research on quantum spin ladders continues to attract considerable
attention from both theoretical and experimental points of view due to
their relevance to a large number of low-dimensional materials, such
as particular cuprates and organic compounds
\cite{exp1,exp2,exp3,exp4,exp5,exp6}, among others.  Initially, most
of the theoretical results concerning ladder systems were obtained
from the standard Heisenberg ladder. Subsequently, other models with
generalized interactions have been proposed.  In this context,
Nersesyan and Tsvelik \cite{ladd1} introduced a  spin ladder model
 incorporating a biquadratic spin exchange interaction term, which, when 
sufficiently strong, exhibits new dimerized phases \cite{ladd2}. 
Various ladder models have been developed by an
extension of the symmetry algebra
\cite{ladd3,ladd4,ladd5,ladd6,ladd7}. A special case of the
Nersesyan-Tsvelik model \cite{ladd1} was proposed later by Wang
\cite{ladd6}. This model, based on the $SU(4)$ symmetry algebra, is
exactly solvable by Bethe ansatz methods and  exhibits a spin gap in
the spectrum of elementary triplet excitations, a necessary condition
for superconductivity to occur under hole doping.  In addition, it was
recently observed \cite{BGFZ} that this model can be used to describe
some physical properties of different types of two-leg ladder
compounds, such as Cu$_{2}$(C$_5$H$_{12}$N$_2$)$_2$Cl$_4$ \cite{exp3},
(C$_5$H$_{12}$N)$_2$CuBr$_{4}$ \cite{exp4}, (5IAP)$_2$CuBr$_4\cdot
2$H$_2$O \cite{exp5} and KCuCl$_3$, TlCuCl$_3$ \cite{exp6}. In the
absence of a magnetic field the model exhibits three quantum phases,
while in the presence of a strong magnetic field there is a gapped
phase in the regime $H < H_{c1}$, a fully polarized gapped phase for
$H > H_{c2}$ and a Luttinger liquid magnetic phase in the regime
$H_{c1} < H < H_{c2}$.  This observation suggests that the physical properties of the  ladder
compounds can be accessed  via the
well-established knowledge of integrable systems.

On the other hand, the effect of boundary impurities and
defects also plays an important role in quasi-one dimensional 
systems. An integrable $SU(4)$ spin ladder model 
with a boundary defect has been proposed and investigated 
recently through Bethe ansatz methods \cite{wang}.
This model, however, is just a particular case
of a more general family of exactly solvable ladder models 
based on the $SU(4)$ symmetry algebra that
can be constructed from more general types of bounday conditions.
Basically, by this strategy, a set of equations to 
deal with the boundaries, called 
reflection equations (RE) are introduced \cite{C,EK}. The solutions of 
these equations \cite{D,op2,op3}, referred to as  $K$-matrices, in turn
introduce boundary interactions into the Hamiltonian of the 
system, in such a manner that integrability is preserved. 
The boundary interaction terms in spin ladder models may be realized by
impurity doping at the ends of the ladder. Impurity doping in a spin
ladder system with a spin gap has been performed \cite{imp-dop}.
Substantial change in macroscopic properties such as enhancement in
spin correlations and magnetic susceptibilities are observed in the
low impurity concentration region.  The boundary impurity doping may
change the critical behaviour at the boundaries of the ladder
systems.

The purpose of this paper is to present a complete family of 
integrable spin ladder systems based on the $SU(4)$ symmetry
algebra with boundary impurities in a systematic way. 
An analytic analysis of the thermodynamic properties 
of these models is then  performed by 
means of the  thermodynamic Bethe ansatz (TBA)
method and, in particular, the effect of these impurities on 
the free energy, the susceptibility and  the magnetization
is discussed. So far, 
the results obtained provide a clear
interpretation of the impurity effects in the low
temperature regime of an integrable open spin ladder system.

The paper is organized as follows. In section \ref{sec2}, we present
the SU(4) solution of the YBE and solve the corresponding
RE. Furthermore, we give the explicit expressions for the Hamiltonian
with different types of boundary defects. Section \ref{sec3} is devoted
to the derivation of the Bethe-ansatz solution by means of the
Quantum Inverse Scattering Method (QISM). The reader more interested in the physics of the model  may choose to skip section \ref{sec3}.  In section \ref{sec4}, the ground state properties, the quantum
phase diagram and the boundary impurity effects are studied via the
TBA. A summary and discussion of our main results  is given  in section \ref{sec5}.

\section{The integrable
spin ladder model with boundary impurities} 
\label{sec2}

Let us begin by
introducing the  integrable spin ladder model based on the
$SU(4)$ symmetry with boundary fields,
\begin{equation} H=\frac{J_{\parallel}}{\gamma}H_{{\rm
leg}}+J_{\perp}\sum_{j=1}^{L}\vec{S}_j\cdot\vec{T}_j
+H_{1}^{(m)}+H_{L}^{(l)},\label{Ham} \end{equation} where the leg part
consists of Heisenberg exchange and four-spin interaction terms
\begin{equation} H_{{\rm
leg}}=\sum_{j=1}^{L-1}\left(\frac{1}{4}+\vec{S}_j\cdot
\vec{S}_{j+1}+\vec{T}_j\cdot
\vec{T}_{j+1}+4\vec{S}_j\cdot\vec{S}_{j+1}\cdot\vec{T}_j\cdot\vec{T}_{j+1}\right).\label{leg-int}
\end{equation} 
The left (right) boundary terms $H_1^{(m)}$
($H_L^{(l)}$) depend on arbitrary parameters
$U_{\pm}$ and are given by 
\begin{eqnarray} 
H_{1}^{(m)} &=&
\left\{\begin{array}{ll} -U_-\vec{S }_1\cdot\vec{T }_1-\frac{1}{4}U_-,
& {\rm for}~~m=1\\
-U_-(\frac{1}{2}-S^z_1)(\frac{1}{2}-T_1^z)+\frac{1}{2}U_-, &
{\rm for}~~m=2 \\ U_-\left(\vec{S}_1\cdot \vec{T
}_1-(\frac{1}{2}-S^z_1)(\frac{1}{2}-T_1^z)\right)+\frac{1}{4}U_-, &
{\rm for}~~m=3\\ U_-\left(\vec{S}_1\cdot \vec{T
}_1-(\frac{1}{2}+S^z_1)(\frac{1}{2}+T_1^z)\right)+\frac{1}{4}U_-, &
{\rm for}~~m=4\\
 -2U_-S_1^zT_1^z+\frac{1}{2}U_-, &{\rm for}~~ m =5
\end{array}\right.,\label{Ham-bt-1}\\ 
H_{L}^{(l)} &=&
\left\{\begin{array}{ll} -U_+\vec{S }_L\cdot\vec{T }_L-\frac{1}{4}U_+,
&  {\rm for} ~~l=1\\
-U_+(\frac{1}{2}-S^z_L)(\frac{1}{2}-T_L^z)+\frac{1}{2}U_+, &
{\rm for}~~l=2\\ U_+\left(\vec{S}_L\cdot\vec{T
}_L-(\frac{1}{2}-S^z_L)(\frac{1}{2}-T_L^z)\right)+\frac{1}{4}U_+, &
{\rm for}~~l=3\\ U_+\left(\vec{S}_L\cdot\vec{T
}_L-(\frac{1}{2}+S^z_L)(\frac{1}{2}+T_L^z)\right)+\frac{1}{4}U_+, &
{\rm for}~~l=4\\ -2U_+S_L^zT_L^z+\frac{1}{2}U_+, &{\rm  for}~~ l=5
\end{array}\right..\label{Ham-bt-2} 
\end{eqnarray} 
In the above $\vec{S
}_j$ and $\vec{T}_j$ are the standard spin$-\frac{1}{2}$ operators
acting on  site $j$ of the upper and lower legs, respectively,
$J_{\parallel}$ and $J_{\perp}$ are the intrachain and interchain
couplings (see figure 1) and $L$ is the length of the ladder. It is
worth noticing that $\gamma$ is a rescaling constant which can  be used to minimize the biquadratic term such that the quantum phase of the model (\ref{Ham}) lies in the same Haldane spin liquid phase as that of
the conventional spin ladder (see section \ref{sec4}).
\begin{figure}[t]
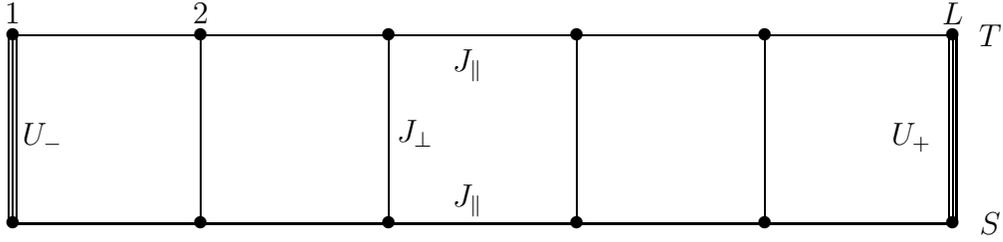

\begin{center} 
\Lattice \\
\end{center}
\caption{The $su(4)$ spin ladder with boundary impurities. $U_{\pm}$ are the boundary impurity coupling constants. $J_{\parallel}$ and $J_{\perp}$ are the intrachain and interchain couplings. }
\end{figure}

Notice that the boundary terms corresponding to the
first case ($m=1,\, l=1$) in (\ref{Ham-bt-1}) and (\ref{Ham-bt-2}) act
as Heisenberg-type rung couplings, whereas in the second case
($m=2$, $l=2$) they act as a $z$-component spin interaction  with
boundary magnetic fields. In the third ($m=3$,
$l=3$) and fourth ($m=4,\, l=4$) cases, they act as a combination of
Heisenberg-type rung coupling and  $z$-component spin interaction
with  boundary magnetic fields.  In the last case ($m=5, \, l=5$) only the $z$-component spin interaction terms survive. 
The Hamiltonian (\ref{Ham}) thus 
contains five different types of boundary rung interactions at each
edge of the ladder realizing different impurity dopings. This leads to
twenty five possible choices of  boundary impurities. The
rung interaction in the bulk was, as usual, introduced by the chemical
potential terms given by $-J_{\perp}\sum ^{L}_{j=1}{e_j}^{11}$ in the
canonical basis $e_i\otimes e_j$. The rung states split into a singlet
and a triplet denoted by 
\begin{eqnarray} |1\rangle & =
&\frac{1}{\sqrt{2}}\left(|\uparrow \downarrow\rangle -|\downarrow
\uparrow \rangle \right),\,\,\, |2\rangle
=|\uparrow\uparrow\rangle,\nonumber\\ |3\rangle & =
&\frac{1}{\sqrt{2}}\left(|\uparrow \downarrow\rangle +|\downarrow
\uparrow \rangle \right),\,\,\,|4\rangle
=|\downarrow\downarrow\rangle,\label{basis} \end{eqnarray}
respectively. The leg interaction part of the Hamiltonian
(\ref{leg-int})
 does not change under the basis transformation (\ref{basis}). However, the bulk rung interaction part and the boundary rung interaction terms
alter with respect to the choice of the order of singlet and triplet in  the  basis (\ref{basis}).

In order to derive this model let us begin by recalling 
 the $SU(4)$ $R$-matrix 
\begin{equation}
\hspace*{-\mathindent}
R_{12}(u)  =
\mbox{\begin{footnotesize}$\left(\matrix{
w_1&0&0&0&0&0&0&0&0&0&0&0&0&0&0&0\cr
0&w_2&0&0&w_3&0&0&0&0&0&0&0&0&0&0&0\cr
0&0&w_2&0&0&0&0&0&w_3&0&0&0&0&0&0&0\cr
0&0&0&w_2&0&0&0&0&0&0&0&0&w_{3}&0&0&0\cr
0&w_3&0&0&w_2&0&0&0&0&0&0&0&0&0&0&0\cr
0&0&0&0&0&w_1&0&0&0&0&0&0&0&0&0&0\cr
0&0&0&0&0&0&w_2&0&0&w_3&0&0&0&0&0&0\cr
0&0&0&0&0&0&0&w_2&0&0&0&0&0&w_3&0&0\cr
0&0&w_3&0&0&0&0&0&w_2&0&0&0&0&0&0&0\cr
0&0&0&0&0&0&w_3&0&0&w_2&0&0&&0&0&0\cr
0&0&0&0&0&0&0&0&0&0&w_1&0&0&0&0&0\cr
0&0&0&0&0&0&0&0&0&0&0&w_2&0&0&w_3&0\cr
0&0&0&w_3&0&0&0&0&0&0&0&0&w_2&0&0&0\cr
0&0&0&0&0&0&0&w_3&0&0&0&0&0&w_2&0&0\cr
0&0&0&0&0&0&0&0&0&0&0&w_3&0&0&w_2&0\cr
0&0&0&0&0&0&0&0&0&0&0&0&0&0&0&w_1}\right)$\end{footnotesize}}
\nonumber\\* \label{R12}
\end{equation}
where the elements are given by 
\begin{equation}
w_1  = u+1,\qquad
w_2  =u,\qquad
w_3  = 1.\label{eq:w}
\end{equation}
The quantum $R$-matrix (\ref{R12}) satisfies the Yang-Baxter equation (YBE)
\begin{equation}
R_{12}(u-v)R_{13}(u)R_{23}(v)
= R_{23}(v)R_{13}(u)R_{12}(u-v),
\label{YBE}
\end{equation}
guaranteeing the integrability of the model with periodic BC. 
This  $R$-matrix enjoys the properties 
\begin{equation}
R_{12}(u)R_{21}(-u)=1-u^2,\qquad
R_{12}^{{\rm t}_1{\rm t}_2}(u)=R_{12}(u),
\end{equation}
where superscript ${\rm t}_a$ denotes the transposition in the space
with index $a$.  For other types of boundary conditions, the YBE will
still account for the integrability of the bulk part of the model, but
the boundary terms have to be chosen appropriately in order to
preserve the integrability. In particular, the left and right
reflection matrices, $K_{-}$ and $K_{+}$, respectively, are required to
satisfy the  RE's \cite{EK,op2}
\begin{eqnarray}
\lefteqn{
 R_{12}(u-v)
 \stackrel{1}{K}_{-}\!(u)\,
 R_{21}(u+v)
 \stackrel{2}{K}_{-}\!(v )} \nonumber\\
& = &
 \stackrel{2}{K}_{-}\!(v )\,
 R_{12}(u+v )
 \stackrel{1}{K}_{-}\!(u)\,
 R_{21}(u-v ),  \label{RE1}\\
\lefteqn{
 R_{21}^{{\rm t}_{1}{\rm t}_2}(v -u)
 \stackrel{1}{K}\!\mbox{}_{+}^{{\rm t}_1}(u)\,
 \tilde{R}_{12}(-u-v)\stackrel{2}{K}\!\mbox{}_{+}^{{\rm t}_2}(v )}
 \nonumber \\
& = &
 \stackrel{2}{K}\!\mbox{}_{+}^{{\rm t}_2}(v )\,
 R_{21}^{{\rm t}_1{\rm t}_2}(-u-v)
 \stackrel{1}{K}\!\mbox{}_{+}^{{\rm t}_1}\!(u)\,
 R_{12}^{{\rm t}_1{\rm t}_2}(v -u).
\label{RE2}
\end{eqnarray}
In the above we have introduced the object $\tilde{R}$  which may be determined by the relations
\begin{equation}
\tilde{R}^{{\rm t}_2}_{12}(-u)R^{{\rm t}_1}_{21}(u)=1,\,\,\,\,\,
\tilde{R}^{{\rm t}_1}_{21}(-u)R^{{\rm t}_2}_{12}(u)=1,
\end{equation}
and we have used the conventional notation
\begin{equation}
\stackrel{1}{X} \equiv X\otimes\mathbf{I}_{V_2},\qquad
\stackrel{2}{X} \equiv \mathbf{I}_{V_1}\otimes X,
\end{equation}
where $\mathbf{I}_{V}$ denotes the identity operator on $V$
and, as usual, $R_{21}={\bf P}\cdot R_{12}\cdot {\bf P}$,  with 
 ${\bf P}$ being the permutation operator. 
After a lengthy calculation we find the  possible solutions of the RE's for
the diagonal  $K_{\pm}$-matrices (see also Ref. \cite{D})
\begin{equation}
K_{\pm}(u)  = 
\left(\matrix{K1_{\pm}(u)&0&0&0\cr
              0&K2_{\pm}(u)&0&0\cr
              0&0&K3_{\pm}(u)&0\cr
              0&0&0&K4_{\pm}(u)\cr}\right). \label{K-PM} 
\end{equation}
The solutions for $K_-$, corresponding to the left boundary are:\\

Case $1$
\begin{equation}
K1_-(u) = u+\xi _-,\,\,
K2_-(u) = K3_-(u)=K4_-(u)=-u+\xi _-;\label{Km1}
\end{equation}
Case $2$
\begin{equation}
K1_-(u) = K2_-(u) = K3_-(u)= u+\xi _-,\,\,
K4_-(u)=-u+\xi _-;\label{Km2}
\end{equation}
Case $3$
\begin{equation}
K1_-(u) = K4_-(u)=-u+\xi _-,\,\,
K2_-(u) = K3_-(u)=u+\xi _-;\label{Km3}
\end{equation}
Case $4$
\begin{equation}
K1_-(u) = K2_-(u)=-u+\xi _-,\,\,
K3_-(u) = K4_-(u)=u+\xi _-;\label{Km4}
\end{equation}
Case $5$
\begin{equation}
K1_-(u) =K3_-(u)=u+\xi _-,\,\,
K2_-(u) = K4_-(u)=-u+\xi _-.\label{Km5}
\end{equation}
On the other hand, the solutions for $K_+$, corresponding to the right boundary, are:  \\

Case $1$
\begin{eqnarray}
K1_+(u) &=& -u+\xi _++3,\nonumber\\
K2_+(u)&  = & K3_+(u)=K4_+(u)=u-\xi _++1;\label{Kp1}
\end{eqnarray}
Case $2$
\begin{eqnarray}
K1_+(u) &=&K2_+(u)  =  K3_+(u) =-u-\xi _+-1,\nonumber\\
K4_+(u)& = &u-\xi _++3;\label{Kp2}
\end{eqnarray}
Case $3$
\begin{eqnarray}
K1_+(u) &=& K4_+(u)=u-\xi _++2,\nonumber\\
K2_+(u) & = & K3_+(u)=-u-\xi _+-2;\label{Kp3}
\end{eqnarray}
Case $4$
\begin{eqnarray}
K1_+(u) &=& K2_+(u)=u-\xi _++2,\nonumber\\
K3_+(u) &=& K4_+(u)=-u-\xi _+-2;\label{Kp4}
\end{eqnarray}
Case $5$
\begin{eqnarray}
K1_+(u) &=& K3_+(u)=-u-\xi _+-2,\nonumber\\
K2_+(u)& = &K4_+(u)=u-\xi _++2.
\label{Kp5}
\end{eqnarray}
In the above $\xi_{\pm}=\frac{J_{\parallel}}{\gamma U_{\pm}}$ are free parameters related to the left $-$ (right $+$) boundary coupling $U_-$ ($U_+$) respectively.
The boundary
pairs $K^{(m)}_{-}(u),\,K_{+}^{(l)}(u)$, $l,m=1,\ldots,5$, 
lead to  the boundary pair terms $H^{(m)}_-$ and $H^{(l)}_+$ in  (\ref{Ham-bt-1}) and (\ref{Ham-bt-2}), whose combination leads to twenty five possible choices of boundary impurities.  A special choice of the  boundary term $H^{(1)}$ was already investigated in  \cite{wang}.  The symmetries enjoyed by the
$R$-matrix (\ref{R12})  and the RE's (\ref{RE1}) and (\ref{RE2}) constitute the
necessary ingredients for the integrability of the model with boundary
impurities, due to the fact that the double-row transfer matrix of the
system
\begin{equation}
\tau (u)= {\rm tr}_0\left[{K}_+(u)T(u)K_-(u)T^{-1}(-u)\right],\label{TM}
\end{equation}
commutes for different values of the spectral parameter.
Here $T(u)$ denotes the monodromy matrix given by 
\begin{equation}
T(u )= 
R_{0,L}(u)R_{0,L-1}(u )\cdots R_{0,2}(u)R_{0,1}(u)
\end{equation}
and $T^{-1}$ its inverse. 
The Hamiltonian (\ref{Ham}) associated with the quantum $R$-matrix
(\ref{R12}) is related to the double-row transfer matrix (\ref{TM})
by
\begin{equation}
H=-\frac{J_{\parallel}}{2\gamma}\frac{d}{du}\ln\tau (u)|_{u=0}-J_{\perp}\sum_{j=1}^{L}e_j^{11}+{\rm const.}
 \label{HTR}
\end{equation}
Here   
\begin{equation}
\frac{d}{du}\ln\tau (u)|_{u=0}=2\sum_{j=1}^{L-1}H_{jj+1}+K_-^{-1}(0)K^{'}_-(0)+
2\frac{{\rm tr_0}K_+(0)R_{0L}^{'}(0)P_{0L}}{{\rm tr_0}K_+(0)},
\end{equation}
where the prime denotes the derivative with respect to the spectral
 parameter. The relation (\ref{HTR}) clearly indicates the
 identification $U_{\pm}=\frac{J_{\parallel}}{\gamma \xi_{\pm}}$ between
 the boundary impurity couplings $U_{\pm}$ and the free parameters
 $\xi_{\pm}$ of the boundary scattering matrices.  So far, we have completed
 the
 first step towards the 
 solution of  the model with boundary impurities. Next we 
 proceed with the diagonalization of the transfer matrix (\ref{TM}) by
 means of the open algebraic Bethe ansatz \cite{ggrsf,G1}.

\section{The algebraic Bethe-ansatz approach}
\label{sec3}
\subsection{First-level nesting structure}
\label{subsec1}

In order to find the spectrum of our Hamiltonian 
with boundary defects, we first need to  solve the eigenvalue problem of the transfer matrix, namely $\tau \Phi=\lambda \Phi$. As usual, the transfer matrix (\ref{TM}) can be written in the  form
\begin{equation}
\tau (\lambda)={\rm tr}_0\left[ K_+(u)\tilde{T}_-(u)\right],
\label{TM2}
\end{equation}
where $\tilde{T}_-(u)$ is the double-row monodromy matrix  defined by
\begin{equation}
\tilde{T}_-(u)=
T(u)K_-(u)T^{-1}(-u).
\end{equation}
One can verify that  $\tilde{T}_-(u)$ also
satisfies the RE (\ref{RE1}).
Following the notation used in Refs.\ \cite{FK,Mar,ggrsf,G1}, we label the elements of the
monodromy matrix $T(u)$ by
\begin{eqnarray}
T(u) & = &
\left(\matrix{A(u)&B_1(u)&B_2(u)&B_3(u)\cr
C_1(u)&D_{11}(u)&D_{12}(u)&D_{13}(u)\cr
C_2(u)&D_{21}(u)&D_{22}(u)&D_{23}(u)\cr
C_3(u)&D_{31}(u)&D_{32}(u)&D_{33}(u)\cr }\right),
\end{eqnarray}
and further
\begin{eqnarray}
T^{-1}(-u) & = &
\left(\matrix{\bar{A}(u)&
\bar{B}_1(u)&\bar{B}_2(u)&\bar{B}_3(u)\cr
\bar{C}_1(u)&\bar{D}_{11}(u)&\bar{D}_{12}(u)&
\bar{D}_{13}(u)\cr
\bar{C}_2(u)&\bar{D}_{21}(u)&\bar{D}_{22}(u)&
\bar{D}_{23}(u)\cr \bar{
C}_3(u)&\bar{D}_{31}(u)&\bar{D}_{32}(u)&
\bar{D}_{33}(u)\cr }
\right), \label{TIM} \\[1ex]
\tilde{T}_-(u) & = &  \left(\matrix{\tilde{B}(u)&
\tilde{A}_1(u)&\tilde{B}_2(u)&\tilde{B}_3(u)\cr
\tilde{C}_1(u)&\tilde{D}_{11}(u)&\tilde{D}_{12}(u)&
\tilde{D}_{13}(u)\cr
\tilde{C}_2(u)&\tilde{D}_{21}(u)&\tilde{D}_{22}(u)&
\tilde{D}_{23}(u)\cr
\tilde{C}_3(u)&\tilde{D}_{31}(u)&\tilde{D}_{32}(u)&
\tilde{D}_{33}(u)\cr }\right).\label{UM}
\end{eqnarray}
According to the first level Bethe ansatz, the eigenvectors $|\Phi\rangle$ of the transfer matrix can be written as
\begin{eqnarray}
|\Phi \rangle =\tilde{B}_{i_1}(u_1)\cdots \tilde{B}_{i_N}(u_{N})|\phi \rangle F_{(1)}^{i_1\cdots i_N}, \label{n-state}
\end{eqnarray}
where the summation is taken on the repeated indicies. The coefficients with indices $i_n=1,2,3,\, n=1,\cdots  ,N$ will be determined later by the second level Bethe ansatz. The first level pseudovacuum $|\phi \rangle $ is chosen as the standard ferromagnetic state 
\begin{equation}
|\phi \rangle =|0\rangle _L\otimes \cdots\otimes|0\rangle _i\otimes\cdots
\otimes |0\rangle _1,\label{pseudo-vacuum}
\end{equation}
where $ |0\rangle _i= \left(1,0,0,0\right)_i^{{\rm t}}$ acts
as a highest-weight vector. This state corresponds to the product of
the rung singlet state in the basis (\ref{basis}). Different choices
of the order of the basis (\ref{basis}) will change the eigenvalues of
the Hamiltonian which facilitates the   analysis of the ground
state in different regions.
From the structure of the $R$-matrix (\ref{R12}), one can deduce that
the operators $B_i(u)$ and $\bar{B}_i(u)$ ($i=1,2,3$) act on the reference state as creation operators  creating particles with
pseudo-momenta $u$ and $-u$, respectively.  The
operators $C_i(u)$ ($i=1,\cdots,3$) behave as
annihilation fields.  Furthermore, using an invariant version  of the
Yang-Baxter algebra,
\begin{equation}
\stackrel{2}{T}\mbox{$\!$}^{-1}(-u)
 R_{12}(2u)
 \stackrel{1}{T}\!(u) =
 \stackrel{1}{T}\!(u)
 R_{12}(2u)
 \stackrel{2}{T}\mbox{$\!$}^{-1}(-u),\label{YBA}
\end{equation}
we obtain, apart from an overall factor
$Q(u)=K1_{-}(u)K1_{+}(u)$, the eigenvalue
of the transfer matrix acting on the reference state $|\phi \rangle$
\begin{equation}
\tau(u)|\phi \rangle  = \left\{ \omega^{+}_{A}(u)\tilde{A}(u)
+\sum_{a=1}^{3}\omega ^{+}_{a}(u)\hat{D}_{aa}(u)\right\}|\phi \rangle,
\label{fact}
\end{equation}
where we have introduced the transformations
\begin{equation}
\hat{D}_{ij}(u)  = 
\tilde{D}_{ij}(u)-\delta_{ij}\frac{w_3(2u)}{w_1(2u)}\tilde{A}(u) = \omega_i^{-}(u)D_{ii}(u)\bar{D}_{ii}(u), \label{transform}
\end{equation}
with $i=1,2,3$ and 
\begin{equation}
\tilde{A}(u)=\omega_{A}^-(u)A(u)\bar{A}(u).
\end{equation}
In the  above expression, 
\begin{eqnarray}
& &
\begin{array}{ll}
\omega_A^-(u)=1, & {\rm for}~~m=1,\ldots ,5,
\end{array}\nonumber \\
& &
\omega_A^+(u)=\left\{\begin{array}{ll}
\frac{(u+2)(u+\xi_+)}{(u+1/2)(u+\xi _++3)},& {\rm for}~~l=1,\\
\frac{(u+2)(u+\xi_+)}{(u+1/2)(u+\xi _++1)},& {\rm for}~~l=2,\\
\frac{(u+2)(-u+\xi_+)}{(u+1/2)(-u+\xi _+-2)},& {\rm for}~~l=3,4,\\
\frac{(u+2)(u+\xi_+)}{(u+1/2)(u+\xi _++2)},& {\rm for}~~l=5,
\end{array}\right.\nonumber\\
& &
\omega_1^-(u)=\left\{\begin{array}{ll}
\frac{u(-u+\xi_--1)}{(u+1/2)(u+\xi _-)},& {\rm for}~~m=1,\\
\frac{u(u+\xi_-+1)}{(u+1/2)(-u+\xi _-)},& {\rm for}~~m=3,\\
\frac{u}{u+1/2},& {\rm for}~~m=2,4,\\
\frac{u(-u+\xi_--1)}{(u+1/2)(u+\xi _-)},& {\rm for}~~m=5,
\end{array}\right.\nonumber\\
& &
\omega_2^-(u)=\left\{\begin{array}{ll}
\frac{u(-u+\xi_--1)}{(u+1/2)(u+\xi _-)},& {\rm for}~~m=1,\\
\frac{u(u+\xi_-+1)}{(u+1/2)(-u+\xi _-)},& {\rm for}~~m=3,\\
\frac{u(u+\xi_-+1)}{(u+1/2)(-u+\xi _-)}  ,& {\rm for}~~m=4,\\
\frac{u}{u+1/2},& {\rm for}~~m=2,5,
\end{array}\right.\nonumber\\
& &
\omega_3^-(u)=\left\{\begin{array}{ll}
\frac{u(-u+\xi_--1)}{(u+1/2)(u+\xi _-)},& {\rm for}~~m=1,\\
\frac{u}{u+1/2},& {\rm for}~~m=3,\\
\frac{u(u+\xi_-+1)}{(u+1/2)(-u+\xi _-)},& {\rm for}~~m=4,\\
\frac{u(-u+\xi_--1)}{(u+1/2)(u+\xi _-)} ,& {\rm for}~~m=2,5,
\end{array}\right.
\nonumber
\end{eqnarray}
with $ \omega_1^+(u)=K2_+(u),\,\omega_2^+(u)=K3_+(u),\,\omega_3^+(u)=K4_+(u)$
 for $l=1, \ldots,5$. In addition, 
$$
A(u)\bar{A}(u)|0\rangle = w_1(u)^2|0\rangle,\,\,\, 
D_{ii}(u)\bar{D}_{ii}(u)|0\rangle=w_2(u)^2|0\rangle.
$$
We note that the operators
$\tilde{B}_i(u)$, $i=1,2,3$, constitute a three-component vector with
both positive and negative pseudo-momenta still playing the role of the
creation fields acting on the pseudovacuum state. In order to make further progress we return to the  RE
(\ref{RE1}) and derive the  commutation relations
\begin{eqnarray}
& &\tilde{A}(u_1)\tilde{B}_a(u_2) = 
\frac{(u_2-u_1+1)(u_1+u_2)}{(u_2-u_1)(u_1+u_2+1)}\tilde{B}_a(u_2)\tilde{A}(u_1)
\nonumber\\
& &
-\frac{(u_1+u_2)}{(u_2-u_1)(u_1+u_2+1)}\tilde{B}_a(u_1)\tilde{A}(u_2)\label{com1}\\
& &-\frac{1}{u_1+u_2+1}\left[\sum_{b=1}^{3}\tilde{B}_b(u_1)\hat{D}_{ba}(u_2)+\delta_{ab}\frac{1}{2u_2+1}\tilde{B}_b(u_1)A(u_2)\right],\nonumber \\
& &\hat{D}_{bd}(u_1)\tilde{B}_c(u_2) =  \frac{(u_1-u_2+1)(u_1+u_2+2)}{(u_1-u_2)(u_1+u_2+1)}\nonumber\\
& &\times \left\{r^{(1)}(u_1+u_2+1)^{eb}_{gh}r^{(1)}(u_1-u_2)^{ih}_{cd}\tilde{B}_e(u_2)\hat{D}_{gi}(u_1)\right\} \nonumber\\
& &-\frac{2(u_1+1)}{(2u_1+1)(u_1-u_2)}r^{(1)}(2u_1+1)^{gb}_{id}\tilde{B}_g(u_1)\hat{D}_{ic}(u_2)\label{com2}\\
& &+\frac{4(u_1+1)u_2}{(2u_1+1)(2u_2+1)(u_1+u_2+1)}r^{(1)}(2u_1+1)^{gb}_{cd}\tilde{B}_g(u_1)\tilde{A}(u_2),\nonumber
\end{eqnarray} 
 between the diagonal fields and the creation fields.
The summation convention is implied for
repeated indices.  The matrix $r^{(1)}$, which satisfies the Yang-Baxter equation, takes the form 
\begin{eqnarray}
& &r^{aa}_{aa}=1,\,a=1,2,3,\,\,\, r^{ab}_{ab}=\frac{1}{u+1},\,a\neq b=1,2,3\nonumber\\
& &r^{ab}_{ba}=\frac{u}{u+1},\,a\neq b=1,2,3. \label{r1}
\end{eqnarray}
We notice that the first term in the rhs of each of the commutation relations
(\ref{com1})--(\ref{com2}) contribute to the eigenvalues of the
transfer matrix,  which should be analytic functions of the spectral
parameter $u$.  Consequently, the residues at singular points must
vanish. This yields the Bethe ansatz equations, which in turn assure
the cancellation of the unwanted terms in the eigenvalues of the
transfer matrix after the whole nesting procedure.
For convenience, we make  a shift in the spectral parameters, $u=v-1/2,~u_i=v_i-1/2$, such that  the eigenvalue of the transfer matrix (\ref{TM}) can be obtained as
\begin{eqnarray}
& &\tau(v)|\Phi\rangle = \Lambda (v,\{v_i\})|\Phi\rangle\nonumber\\
& & = W_A^{+}(v-1/2)W^-_A(v-1/2)\prod _{i=1}^{N}\frac{(v-v_i-1)(v+v_i-1)}{(v-v_i)(v+v_i)}|\Phi\rangle \label{eign-1}\\
& &
+W^+_a(v-1/2)W_a^-(v-1/2)\prod^{N}_{i=1}\frac{(v-v_i+1)(v+v_i+1)}{(v-v_i)(v+v_i)}\Lambda ^{(1)}(v,\{v_i\})|\Phi\rangle ,\nonumber
\end{eqnarray}
provided that 
\begin{eqnarray}
\frac{W^+_A(v_i-1/2)W^-_A(v_i-1/2)(2v_i-1)}{W_1^+(v_i-1/2)W^-_1(v_i-1/2)(2v_i+1)}\nonumber\\
=\prod^{N}_{\stackrel{\scriptstyle l=1}{l\neq i}}\frac{(v_i-v_l+1)(v_i+v_l+1)}{(v_i-v_l-1)(v_i+v_l-1)}\Lambda^{(1)} (v,\{v_i\})\mid _{v=v_i}.
\end{eqnarray}
Here $a=1,2,3$, and we appropriately choose  
\begin{eqnarray}
& &
\begin{array}{ll}
W_A^-(u)=1, & {\rm for}~~m=1,\ldots ,5,\label{W1}
\end{array}\\
& &
W_A^+(u)=\left\{\begin{array}{ll}
\frac{(u+2)(u+\xi_+)}{(u+1/2)(u+\xi _++3)},& {\rm for}~~l=1,\\
\frac{(u+2)(u+\xi_+)}{(u+1/2)(u+\xi _++1)},& {\rm for}~~l=2,\\
\frac{(u+2)(-u+\xi_+)}{(u+1/2)(-u+\xi _+-2)},& {\rm for}~~l=3,4,\\
\frac{(u+2)(u+\xi_+)}{(u+1/2)(u+\xi _++2)},& {\rm for}~~l=5,
\end{array}\right.\\
& &
W_a^+(u)=\left\{ \begin{array}{ll}
\frac{u-\xi_++1}{-u-\xi_+-3}, & {\rm for}~~l=1,\\
1, & {\rm for}~~l=2,\ldots ,5,
\end{array}\right.\\
& &
W_a^-(u)=\left\{ \begin{array}{ll}
\frac{u(-u+\xi_--1)}{(u+1/2)(u+\xi _-)},& {\rm for}~~m=1,\\
\frac{u}{u+1/2}, & {\rm for}~~m=2,\ldots ,5.\label{W3}
\end{array}\right.
\end{eqnarray}
$\Lambda ^{(1)}(v,\{v_i\})$ is the eigenvalue of the second level transfer matrix $\tau ^{(1)}$ related to an $SU(3)$ invariant open chain, i.e.
\begin{equation}
\tau ^{(1)} = Tr_0K_+^{(1)}(v)T^{(1)}(v)K_-^{(1)}(v)\bar{T}^{(1)}(v),\label{TM-1}
\end{equation}
where 
\begin{eqnarray}
T^{(1)}(v,\{v_i\}) & = &
 r_{12}^{(1)}(v+v_1)_{h_{1}g_{1}}^{e_{1}a}\ldots
 r_{12}^{(1)}(v+v_N)_{h_{N}g_{N}}^{e_{N}g_{N-1}}, \\
\bar{T}^{(1)}(v,\{v_i\}) &= &
 r_{21}^{(1)}(v-v_N)_{l_{N-1}i_{N}}^{e_{N}h_{N}}\ldots
 r_{21}^{(1)}(v-v_1)_{ai_{1}}^{e_{1}h_{1}}.
\label{Tinv}
\end{eqnarray}
Here  we have  used the standard notation
$r_{12}^{(1)}(v)=P\cdot r^{(1)}(v)$
where $P$ is the  standard permutation operator, which can be represented by
a $3^2\times 3^2$ matrix, i.e., $ p_{\alpha\beta, \gamma
\delta}=\delta_{\alpha\delta}\delta_{\beta\gamma}$.
It can be seen that the coefficients $F_{(1)}^{i_{1},\ldots,i_{n}}$
act as  the multi-particle vectors for the inhomogeneous transfer
matrix (\ref{TM-1}). We remark that the coefficients $W$ given in  (\ref{W1})-(\ref{W3}) are  chosen in order to match the choice of the transfer matrix (\ref{TM-1}) with the  nested $K_{\pm}^{(1)}$-matrices,
\begin{equation}
K_{\pm}^{(1)}(v) = 
\left(\matrix{K1^{(1)}_{\pm}(v)&0&0\cr
              0&K2^{(1)}_{\pm}(v)&0\cr
              0&0&K3^{(1)}_{\pm}(v)\cr
             }\right). \label{KPM-1} 
\end{equation}
Now corresponding to the first solution (\ref{Kp1}), we have
\begin{eqnarray}
K1^{(1)}_-(v) & = &K2^{(1)}_-(v) = K3^{(1)}_-(v) =  1,\label{K1-1}\\
K1^{(1)}_+(v) & = &K2^{(1)}_+(v) = K3^{(1)}_+(v) =  1.
\end{eqnarray}
And to the second solution (\ref{Kp2}):
\begin{eqnarray}
K1^{(1)}_-(v) & = &K2^{(1)}_-(v) = 1,\,\,\,
K3^{(1)}_-(v) =\frac{-v+\xi _--1/2}{v+\xi _--1/2},\\
K1^{(1)}_+(v) & = &K2^{(1)}_+(v) = 1,\,\,\, 
K3^{(1)}_+(v)=\frac{v-\xi _++5/2}{-v-\xi _+-1/2}.
\end{eqnarray}
And to the third solution (\ref{Kp3}):
\begin{eqnarray}
K1^{(1)}_-(v) & = &K2^{(1)}_-(v) = \frac{v+\xi _-+1/2}{-v+\xi _-+1/2},\,\,\,K3^{(1)}_-(v) =  1,\\
K1^{(1)}_+(v) & = &K2^{(1)}_+(v) = \frac{-v-\xi _+-3/2}{v-\xi _++3/2},\,\,\,K3^{(1)}_+(v) =  1.
\end{eqnarray}
And to the fourth solution (\ref{Kp4}):
\begin{eqnarray}
K1^{(1)}_-(v) &=& 1,\,\,\,K2^{(1)}_-(v) = K3^{(1)}_-(v) = \frac{v+\xi _-+1/2}{-v+\xi _-+1/2},\\
K1^{(1)}_+(v) & = & 1,\,\,\,K2^{(1)}_+(v) = K3^{(1)}_+(v) = \frac{-v-\xi _+-3/2}{v-\xi _++3/2}.
\end{eqnarray}
And to the last solution (\ref{Kp5}): 
\begin{eqnarray}
K2^{(1)}_-(v) &=& 1,\,\,\,K1^{(1)}_-(v) = K3^{(1)}_-(v) = \frac{-v+\xi _--1/2}{v+\xi _--1/2},\\
K2^{(1)}_+(v) & = & 1,\,\,\,K1^{(1)}_+(v) = K3^{(1)}_+(v) = \frac{v-\xi _++3/2}{-v-\xi _+-3/2}.\label{K1-5}
\end{eqnarray}
We can show that the reflection matrices (\ref{KPM-1}) with the entries
(\ref{K1-1})-(\ref{K1-5})  do satisfy the
RE
\begin{eqnarray}
\lefteqn{
 r^{(1)}_{12}(u-v)
 \stackrel{1}{K^{(1)}_{-}}\!(u)\,
 r_{21}^{(1)}(u+v)
 \stackrel{2}{K^{(1)}_{-}}\!(v )} \nonumber\\
& = &
 \stackrel{2}{K^{(1)}}_{-}\!(v )\,
 r_{12}^{(1)}(u+v )
 \stackrel{1}{K_{-}^{(1)}}\!(u)\,
 r_{21}^{(1)}(u-v ),  \label{RE1-1}\\
\lefteqn{
 {r_{21}^{(1)}}^{{\rm t}_{1}{\rm t}_2}(v -u)
 {\stackrel{1}{K^{(1)}_{+}}}^{{\rm t}_1}(u)\,
 \tilde{r}_{12}^{(1)}(-u-v){\stackrel{2}{K^{(1)}_{+}}}^{{\rm t}_2}(v )}
 \nonumber \\
& = &
 {\stackrel{2}{K^{(1)}_{+}}}^{{\rm t}_2}(v )\,
 {r_{21}^{(1)}}^{{\rm t}_1{\rm t}_2}(-u-v)
 {\stackrel{1}{K^{(1)}_{+}}}^{{\rm t}_1}\!(u)\,
 {r_{12}^{(1)}}^{{\rm t}_1{\rm t}_2}(v -u).
\label{RE2-1}
\end{eqnarray}
\subsection{Second-level Bethe ansatz}
\label{subsec5}

In order to proceed in the nested algebraic Bethe ansatz, we
have to repeat the whole procedure presented above for the internal block of the monodromy matrix.  Similarly, we
rewrite the second level transfer matrix $\tau ^{(1)} $ (\ref{TM-1}) in the  form
\begin{equation}
\tau ^{(1)}(v)={\rm Tr}_0\left[ K^{(1)}_+(v)U^{(1)}_-(v)\right],
\label{TM2-1}
\end{equation}
where $U^{(1)}_-(v)$ is defined by
\begin{equation}
\tilde{T}^{(1)}_-(v)=
T^{(1)}(v)K^{(1)}_-(v)\bar{T}^{(1)}(v).
\end{equation}
Now we  label the elements of the
monodromy matrices  by
\begin{eqnarray}
T ^{(1)}(\lambda) & = &
\left(\matrix{A ^{(1)}(v)&B_1 ^{(1)}(v)&B_2 ^{(1)}(v)\cr
C_1 ^{(1)}(v)&D_{11} ^{(1)}(v)&D_{12} ^{(1)}(v)\cr
C_2 ^{(1)}(v)&D_{21} ^{(1)}(v)&D_{22} ^{(1)}(v)\cr}\right),\\
\bar{T}^{(1)}(v) & = &
\left(\matrix{\bar{A}^{(1)}(v)&
\bar{B}^{(1)}_1(v)&\bar{B}_2^{(1)}(v)\cr
\bar{C}^{(1)}_1(v)&\bar{D}^{(1)}_{11}(v)&\bar{D}^{(1)}_{12}(v)\cr
\bar{C}^{(1)}_2(v)&\bar{D}^{(1)}_{21}(v)&\bar{D}^{(1)}_{22}(v)\cr }
\right), \label{TIM-1} \\[1ex]
\tilde{T}_-^{(1)}(v) & = &  \left(\matrix{\tilde{A}^{(1)}(v)&
\tilde{B}^{(1)}_1(v)&\tilde{B}^{(1)}_2(v)\cr
\tilde{C}^{(1)}_1(v)&\tilde{D}^{(1)}_{11}(v)&\tilde{D}^{(1)}_{12}(v)\cr
\tilde{C}^{(1)}_2(v)&\tilde{D}^{(1)}_{21}(v)&\tilde{D}^{(1)}_{22}(v)\cr
}\right).\label{UM-1}
\end{eqnarray}
From the structure of the $r^{(1)}$-matrix (\ref{r1}), it is found  that
the operators $B^{(1)}_a(v)$ and $\bar{B}^{(1)}_a(v)$ ($a=1,2$) act as
creation fields acting on the reference state $ |0\rangle _i= \left(\matrix{1\cr 0\cr 0}\right)_i$ . The
operators $C_i^{(1)}(v)$ ($i=1,\cdots,2$) behave as
annihilation fields. In order to make further progress we return to the  RE
(\ref{RE1-1}) and derive commutation relations,
\begin{eqnarray}
& & \tilde{A}^{(1)}(v_1)\tilde{B}^{(1)}_a(v_2)= 
\frac{(v_1-v_2-1)(v_1+v_2)}{(v_1-v_2)(v_1+v_2+1)}\tilde{B}^{(1)}_a(v_2)\tilde{A^{(1)}}(v_1)
\nonumber\\
& &
+\frac{(v_1+v_2)}{(v_1-v_2)(v_1+v_2+1)}\tilde{B}^{(1)}_a(v_1)\tilde{A}^{(1)}(v_2)\label{com1-1}\\
& &-\frac{1}{v_1+v_2+1}\left[\sum_{b=1}^{2}\tilde{B}^{(1)}_b(v_1)\hat{D}^{(1)}_{ba}(v_2)+\delta_{ab}\frac{1}{2v_2+1}\tilde{B}^{(1)}_b(v_1)A^{(1)}(v_2)\right],\nonumber \\
& &\hat{D}^{(1)}_{bd}(v_1)\tilde{B}^{(1)}_c(v_2) =  \frac{(v_1-v_2+1)(v_1+v_2+2)}{(v_1+v_2+1)(v_1-v_2)}\nonumber\\
& &\times \left\{r^{(2)}(v_1+v_2+1)^{eb}_{gh}r^{(2)}(v_1-v_2)^{ih}_{cd}\tilde{B}^{(1)}_e(v_2)\hat{D}_{gi}(v_1)\right\} \nonumber\\
& &-\frac{2(v_1+1)}{(2v_1+1)(v_1-v_2)}r^{(2)}(2v_1+1)^{gb}_{id}\tilde{B}^{(1)}_g(u_1)\hat{D}^{(1)}_{ic}(v_2)\label{com2-1}\\
& &+\frac{4(v_1+1)v_2}{(2v_1+1)(2v_2+1)(v_1+v_2+1)}r^{(1)}(2v_1+1)^{gb}_{cd}\tilde{B}^{(1)}_g(v_1)\tilde{A}^{(1)}(v_2),\nonumber
\end{eqnarray} 
between the diagonal
 and the creation fields.
Where again the summation convention is implied for
the repeated indices.  The matrix $r^{(2)}(v)$  is nothing but the $SU(2)$ invariant $R$-matrix, i.e. 
\begin{eqnarray}
& &r^{aa}_{aa}=1,\,a=1,2,\,\,\, r^{ab}_{ab}=\frac{1}{v+1},\,a\neq b=1,2,\nonumber\\
& &r^{ab}_{ba}=\frac{v}{v+1},\,a\neq b=1,2. \label{r2}
\end{eqnarray}
If we define the second level Bethe ansatz  as 
\begin{equation}
|\Psi ^{(1)}\rangle =\tilde{B}^{(1)}_{l_1}(\mu_1)\cdots \tilde{B}^{(1)}_{l_M}(\mu_{M})|0\rangle {F_{(2)}}^{l_1\cdots l_M},
\end{equation} and make the rescalings $\mu _l\rightarrow \mu _l-1/2$ and $v=\tilde{v}-1/2$
we  obtain from the commutation relations (\ref{com1-1}) and (\ref{com2-1}) the eigenvalue $\Lambda ^{(1)}(v,\{v_i\}\{\mu_j\})$, i.e.
\begin{eqnarray}
& &\tau ^{(1)}(\tilde{v})|\Psi ^{(1)}\rangle =\Lambda ^{(1)}(\tilde{v},\{v_i\}\{\mu_j\})|\Psi ^{(1)}\rangle \nonumber\\
& &
 = \left\{W_{+A}^{(1)}(\tilde{v}-1/2)W^{(1)}_{-A}(\tilde{v}-1/2)\prod _{l=1}^{M}\frac{(\tilde{v}-\mu_l-1)(\tilde{v}+\mu_l-1)}{(\tilde{v}-\mu_l)(\tilde{v}+\mu_l)}\right.\label{eign-2}\\
& &
+\left.W^{(1)}_{+a}(\tilde{v}-1/2)W_{-a}^{(1)}(\tilde{v}-1/2)\prod^{N}_{i=1}\frac{(\tilde{v}-v_i-1/2)(\tilde{v}+v_i-1/2)}{(\tilde{v}-v_i+1/2)(\tilde{v}+v_i+1/2)}\right.\nonumber\\
& &\left. \times \prod _{l=1}^{M}\frac{(\tilde{v}-\mu _l+1)(\tilde{v}+\mu _l+1)}{(\tilde{v}-\mu _l)(\tilde{v}+\mu _l)}
\Lambda ^{(2)}(\tilde{v},\{v_i\},\{\mu _l\})\right\}|\Phi ^{(1)}\rangle, \nonumber
\end{eqnarray}
provided that 
\begin{eqnarray}
& &\frac{W^{(1)}_{+A}(\mu _l-1/2)W^{(1)}_{-A}(\mu _l-1/2)(2\mu _l-1)}{W_{+1}^{(1)}(\mu _l-1/2)W^{(1)}_{-1}(\mu _l-1/2)(2\mu _l+1)}\nonumber\\
& &=\prod _{i=1}^{N}\frac{(\mu_l-v_i-1/2)(\mu _l+v_i-1/2)}{(\mu_l-v_i+1/2)(\mu _l+v_i+1/2)}\nonumber\\
& &\times \prod^{M}_{\stackrel{\scriptstyle i=1}{i\neq l}}\frac{(\mu _l-\mu_i+1)(\mu _l+\mu _i+1)}{(\mu _l-\mu_i-1)(\mu _l+\mu _i-1)}\Lambda^{(2)}(\tilde{v},\{v_i\},\{\mu _l\})\mid _{\tilde{v}=\mu _l}.\label{Bethe-2}
\end{eqnarray}
Here $a=1,2$, and 
\begin{eqnarray}
& &
W_{-A}^{(1)}(v)=\left\{\begin{array}{ll}
\frac{v+\xi _-+1/2}{-v+\xi_-+1/2}, & {\rm for}~~m=3,\\
1,& {\rm for}~~m=1,2,4,\\
\frac{-v+\xi _--1/2}{v+\xi_--1/2}, & {\rm for}~~m=5,
\end{array}\right.\\
& &
W_{+A}^{(1)}(v)=\left\{\begin{array}{ll}
\frac{v+3/2}{v+1/2},& {\rm for}~~l=1,\\
\frac{(v+3/2)(v+\xi_+-1/2)}{(v+1/2)(v+\xi _++1/2)},& {\rm for}~~l=2,\\
\frac{(v+3/2)(v+\xi_++1/2)}{(v+1/2)(-v+\xi _+-3/2)},& {\rm for}~~l=3,\\
\frac{(v+3/2)(-v+\xi_++1/2)}{(v+1/2)(-v+\xi _+-3/2)},& {\rm for}~~l=4,\\
\frac{(v+3/2)(-v+\xi_+-1/2)}{(v+1/2)(v+\xi _++3/2)},& {\rm for}~~l=5,
\end{array}\right.\\
& &
W_{-1}^{(1)}(v)=W_{-2}^{(1)}(v)=\left\{\begin{array}{ll}
\frac{v}{v+1/2},& {\rm for}~~m=1,\\
\frac{v}{(v+1/2)(v+\xi_--1/2)},& {\rm for}~~m=2,\\
\frac{v}{(v+1/2)(-v+\xi_-+1/2)},& {\rm for}~~m=3,\\
\frac{v(v+\xi _-+3/2)}{(v+1/2)(-v+\xi_-+1/2)} ,& {\rm for}~~m=4,\\
\frac{v}{(v+1/2)(v+\xi_--1/2)}, & {\rm for}~~m=5,
\end{array}\right.\\
& &
W_{+1}^{(1)}(v)=W_{+2}^{(1)}(v)=\left\{\begin{array}{ll}
1,& {\rm for}~~l=1,\\
\frac{1}{v+\xi _++1/2},& {\rm for}~~l=2,\\
\frac{1}{-v+\xi _+-3/2},& {\rm for}~~l=3,\\
\frac{v+\xi_++3/2}{-v+\xi _+-3/2},& {\rm for}~~l=4,\\
\frac{1}{v+\xi _++3/2},& {\rm for}~~l=5.
\end{array}\right.
\end{eqnarray}
Now $\Lambda ^{(2)}(\tilde{v},\{v_i\},\{\mu_l\})$ is the eigenvalue of the third level  transfer matrix $\tau ^{(2)}$ related to an $SU(2)$ invariant open chain, i.e.
\begin{equation}
\tau ^{(2)} = Tr_0K_+^{(2)}(\tilde{v})T^{(2)}(\tilde{v})K_-^{(2)}(\tilde{v})\bar{T}^{(2)}(\tilde{v}),\label{TM-2}
\end{equation}
where 
\begin{eqnarray}
T^{(2)}(\tilde{v},\{v_i\},\{\mu_l\}) & = &
 r_{12}^{(2)}(\tilde{v}+\mu_1)_{h_{1}g_{1}}^{e_{1}a}\ldots
 r_{12}^{(2)}(\tilde{v}+\mu_M)_{h_{M}g_{M}}^{e_{M}g_{M-1}}, \\
\bar{T}^{(1)}(\tilde{v},\{v_i\},\{\mu _l\}) &= &
 r_{21}^{(2)}(\tilde{v}-\mu_M)_{l_{M-1}i_{M}}^{e_{M}h_{M}}\ldots
 r_{21}^{(2)}(\tilde{v}-\mu_1)_{ai_{1}}^{e_{1}h_{1}}.
\end{eqnarray}
\subsection{Third-level Bethe ansatz}
\label{subsec6}

It has been  shown  so far that the eigenvalue problem of the 
transfer matrix is reduced to  the diagonalization of the
isotropic Heisenberg  model with  $K$-matrices
\begin{eqnarray}
& &K_{-}^{(2)}(\tilde{v})=\left\{\begin{array}{ll}
\left(\matrix{\tilde{v}+\xi_--1&0\cr
0&-\tilde{v}+\xi _--1\cr }\right),&{\rm for}~~m=2,\\
\left(\matrix{\tilde{v}+\xi_-&0\cr
0&-\tilde{v}+\xi _-\cr }\right),&{\rm for}~~m=3,5,\\
\left(\matrix{1&0\cr
0&1\cr }\right),&{\rm for}~~m=1,4,\end{array}\right.\\
& &
K_{+}^{(2)}(\tilde{v})=\left\{\begin{array}{ll}
\left(\matrix{\tilde{v}+\xi_+&0\cr
0&-\tilde{v}+\xi _+\cr }\right),&{\rm for}~~l=2,\\
\left(\matrix{\tilde{v}+\xi_++1&0\cr
0&-\tilde{v}+\xi _+-1\cr }\right),&{\rm for}~~l=3,5,\\
\left(\matrix{1&0\cr
0&1\cr }\right),&{\rm for}~~l=1,4. \end{array}\right.
\end{eqnarray}
Following  the derivation in \cite{EK}, we immediately obtain the eigenvalues of the
nested transfer matrix (\ref{TM-2}), given by 
\begin{eqnarray}
& &
\tau ^{(2)}(\tilde{v}){ F^{(1)}}^{l_1\cdots l_M}=\Lambda ^{(2)}(v,\{v_i\},\{\mu_l\},\{w_q\}) {F^{(1)}}^{l_1\cdots l_M}\nonumber\\
& &
 = \left\{W_{+1}^{(2)}(\tilde{v})W^{(2)}_{-1}(\tilde{v})\prod _{l=1}^{Q}\frac{(\tilde{v}-w_l-1)(\tilde{v}+w_l)}{(\tilde{v}-w_l)(\tilde{v}+w_l+1)}\right.\label{eign-3}\\
& &
+\left.W^{(2)}_{+2}(\tilde{v})W_{-2}^{(2)}(\tilde{v})\prod^{M}_{l=1}\frac{(\tilde{v}-\mu _l)(\tilde{v}+\mu _l)}{(\tilde{v}-\mu_l+1)(\tilde{v}+\mu _l+1)}\right.\nonumber\\
& &\left. \times \prod _{l=1}^{Q}\frac{(\tilde{v}-w_l+1)(\tilde{v}+w_l+2)}{(\tilde{v}-w_l)(\tilde{v}+w_l+1)}\right\}{F^{(1)}}^{l_1\cdots l_M} ,\nonumber
\end{eqnarray}
provided that 
\begin{eqnarray}
& &\frac{W^{(2)}_{+1}(w_l)W^{(2)}_{-1}(w_l)w_l}{W_{+2}^{(2)}(w_l)W^{(2)}_{-2}(w_l)(w_l+1)}=\prod _{j=1}^{M}\frac{(w_l-\mu_j)(w_l+\mu_j)}{(w_l-\mu _j+1)( w_l+\mu_j+1)}\nonumber\\
& &\times \prod^{Q}_{\stackrel{\scriptstyle m=1}{m\neq l}}\frac{(w_l-w_m+1)(w_l+w_m+2)}{(w_l-w_m-1)(w_l+w_m)}.\label{Bethe-3}
\end{eqnarray}
Here
\begin{eqnarray}
& &
W_{-1}^{(2)}(\tilde{v})=\left\{\begin{array}{ll}
\tilde{v}+\xi _--1, & {\rm for}~~m=2,\\
\tilde{v}+\xi _-, & {\rm for}~~m=3,5,\\
1,& {\rm for}~~m=1,4,
\end{array}\right.\\
& &
W_{+1}^{(2)}(\tilde{v})=\left\{\begin{array}{ll}
\frac{(\tilde{v}+1)(\tilde{v}+\xi_+-1)}{2\tilde{v}+1},& {\rm for}~~l=2,\\
\frac{(\tilde{v}+1)(\tilde{v}+\xi_+)}{2\tilde{v}+1},& {\rm for}~~l=3,5,\\
\frac{\tilde{v}+1}{\tilde{v}+1/2},& {\rm for}~~l=1,4,
\end{array}\right.\\
& &
W_{-2}^{(2)}(\tilde{v})=\left\{\begin{array}{ll}
\frac{-\tilde{v}(\tilde{v}-\xi _-+2)}{\tilde{v}+1/2},& {\rm for}~~m=2,\\
\frac{-\tilde{v}(\tilde{v}-\xi _-+1)}{\tilde{v}+1/2},& {\rm for}~~m=3,5,\\
\frac{\tilde{v}}{\tilde{v}+1/2} ,& {\rm for}~~m=1,4,
\end{array}\right.\\
& &
W_{+2}^{(2)}(\tilde{v})=\left\{\begin{array}{ll}
-\tilde{v}+\xi _+-2,& {\rm for}~~l=2,\\
-\tilde{v}+\xi _+-1,& {\rm for}~~l=3,5,\\
1,& {\rm for}~~l=1,4.
\end{array}\right.
\end{eqnarray}
The eigenvalues 
(\ref{eign-2}) and (\ref{eign-3}) as well as  the constraints (\ref{Bethe-2}) (\ref{Bethe-3}) on the rapidities
$\mu_l$ and $w_l$ have paved the way for the complete  diagonalization of  the transfer matrix
(\ref{TM}). Making a further shift on the rapidities,
 $w_l\rightarrow w_l -1/2$, $w_m\rightarrow w_m -1/2$  and $\tilde{v}=v+1/2$,
the eigenvalues of the transfer matrix (\ref{TM}) are  given by 
\begin{eqnarray}
& &\Lambda (v,\{v_i\},\{\mu _l\},\{w_j\})=K1_-(v-\frac{1}{2})K1_+(v-\frac{1}{2}) \nonumber \\
& &
\left\{W^+_A(v-\frac{1}{2})W^-_A(v-\frac{1}{2})(v+\frac{1}{2})^{2L}\right.\nonumber\\
& &\left.\times \prod^{N}_{i=1}\frac{(v-v_i-1)(v+v_i-1)}{(v-v_i)(v+v_i)}\right.\nonumber\\
& &
+\left.W^+_1(v-\frac{1}{2})W^-_1(v-\frac{1}{2})W^{(1)}_{+A}(v)W^{(1)}_{-A}(v)(v-\frac{1}{2})^{2L}\right.\nonumber\\
& &
\left.\times \prod^{N}_{i=1}\frac{(v-v_i+1)(v+v_i+1)}{(v-v_i)(v+v_i)}
\prod^{M}_{l=1}\frac{(v-u_l-\frac{1}{2})(v+u_l-\frac{1}{2})}{(v-u_l+\frac{1}{2})(v+u_l+\frac{1}{2})}\right.\nonumber\\
& &
\left.+W^+_2(v-\frac{1}{2})W^-_2(v-\frac{1}{2})W^{(1)}_{+1}(v)W^{(1)}_{-1}(v)W^{(2)}_{+1}(v+\frac{1}{2})W^{(2)}_{-1}(v+\frac{1}{2})(v-\frac{1}{2})^{2L}\right.\nonumber\\
& &
\left.\times \prod^{M}_{l=1}\frac{(v-\mu_l+\frac{3}{2})(v+\mu_l+\frac{3}{2})}{(v-\mu _l+\frac{1}{2})(v+\mu _l+\frac{1}{2})}
\prod^{Q}_{j=1}\frac{(v-w_j)(v+w_j)}{(v-w_j+1)(v+w_l+1)}\right.\nonumber\\
& &
\left.+W^+_3(v-\frac{1}{2})W^-_3(v-\frac{1}{2})W^{(1)}_{+2}(v)W^{(1)}_{-2}(v)W^{(2)}_{+2}(v+\frac{1}{2})W^{(2)}_{-2}(v+\frac{1}{2})(v-\frac{1}{2})^{2L}\right. \nonumber\\
& &
\left.\times \prod^{Q}_{j=1}\frac{(v-w_j+2)(v+w_j+2)}{(v-w_j+1)(v+w_l+1)}\right\}. \label{eigen}
\end{eqnarray}
The three rapidities $\left\{ v_i,\mu _j, w_k \right\}$ of flavor waves satisfy the Bethe ansatz equations
\begin{eqnarray}
& &\zeta(v_i,\xi_+)\zeta(v_i,\xi_-)\frac{(v_i+\frac{1}{2})^{2L}}{(v_i-\frac{1}{2})^{2L}}\nonumber\\
& &
=\prod^{N}_{\stackrel{\scriptstyle l=1}{l\neq i}}\frac{(v_i-v_l+1)(v_i+v_l+1)}{(v_i-v_l-1)(v_i+v_l-1)}\prod^{M}_{l=1}\frac{(v_i-\mu_l-\frac{1}{2})(v_i+\mu_l-\frac{1}{2})}{(v_i-\mu_l+\frac{1}{2})(v_i+\mu_l+\frac{1}{2})},\label{Bethe1}\\
& &
\eta (\mu_j,\xi_+)\eta(\mu_j,\xi_-)\prod^{N}_{i=1}\frac{(\mu_j-v_i+\frac{1}{2})(\mu_j+v_i+\frac{1}{2})}{(\mu_j-v_i-\frac{1}{2})(\mu_j+v_i-\frac{1}{2})}\nonumber\\
& &=
\prod^{M}_{\stackrel{\scriptstyle i=1}{i\neq j}}\frac{(\mu_j-\mu_i+1)(\mu_j+\mu_i+1)}{(\mu_j-\mu_i-1)(\mu_j+\mu_i-1)}\prod^{Q}_{l=1}\frac{(\mu_j-w_l-\frac{1}{2})(\mu_j+w_l-\frac{1}{2})}{(\mu_j-w_l+\frac{1}{2})(\mu_j+w_l+\frac{1}{2})}\label{Bethe2}\\
& &
\Omega (w_k,\xi_+)\Omega (w_k,\xi_-)\prod^{M}_{l=1}\frac{(w_k-\mu_l+\frac{1}{2})
(w_k+\mu_l+\frac{1}{2})}{(w_k-\mu_l-\frac{1}{2})(w_k+\mu_l-\frac{1}{2})}\nonumber\\
& &=\prod^{Q}_{\stackrel{\scriptstyle l=1}{l\neq k}}\frac{(w_k-w_l+1)(w_k+w_l+1)}{(w_k-w_l-1)(w_k+w_l-1)},\label{Bethe3}
\end{eqnarray}
for $i=1,\ldots,N$, $j=1,\ldots ,M$ and  $k=1,\ldots ,Q$, respectively. Here, we have introduced the  notation
\begin{eqnarray}
& &\zeta(v_{i},\xi_{\pm}) =
\left\{\begin{array}{ll}
\frac{v_i+\xi_{\pm}-\frac{1}{2}}{v_i-\xi_{\pm}+\frac{1}{2}},& {\rm for}~~l=1,~~~m=1,\\
\frac{v_i-\xi_{\pm}-\frac{1}{2}}{v_i+\xi_{\pm}+\frac{1}{2}},& {\rm for}~~l=3,~~~m=3,\\
1,& {\rm for}~~l=2,4,~~~m=2,4,\\
\frac{v_i+\xi_{\pm}-\frac{1}{2}}{v_i-\xi_{\pm}+\frac{1}{2}},& {\rm for}~~l=5,~~~m=5,
\end{array}\right.\\
& &
\eta(\mu_{j},\xi_{\pm}) =
\left\{\begin{array}{ll}
1,& {\rm for}~~l=1,2,3,~~~m=1,2,3\\
\frac{\mu_j+\xi_{\pm}-\frac{1}{2}}{\mu_j-\xi_{\pm}+\frac{1}{2}},& {\rm for}~~l=4,~~~m=4,\\
\frac{\mu_j-\xi_{\pm}}{\mu_j+\xi_{\pm}},& {\rm for}~~l=5,~~~m=5,
\end{array}\right.\\
& &
\Omega (w_{k},\xi_{\pm}) =
\left\{\begin{array}{ll}
\frac{w_k+\xi_{\pm}-\frac{3}{2}}{w_k-\xi_{\pm}+\frac{3}{2}},& {\rm for}~~l=2,~~~m=2,\\
\frac{w_k+\xi_{\pm}-\frac{1}{2}}{w_k-\xi_{\pm}+\frac{1}{2}},& {\rm for}~~l=3,5,~~~m=3,5,\\
1,& {\rm for}~~l=1,4,~~~m=1,4.
\end{array}\right.
\end{eqnarray}
These boundary factors coupled to the three degrees of freedom will result in  a rich physical scenario. 
From (\ref{HTR}) and  (\ref{eigen}),   we  finally obtain the
eigenvalues of the Hamiltonian (\ref{Ham}) as
\begin{eqnarray}
E  = \left\{\begin{array}{ll}
\frac{U_+}{2}+\frac{U_-}{2}+(\frac{J_{\parallel}}{\gamma}-J_{\perp})L\\
+\sum_{i=1}^{N}\left(\frac{J_{\parallel}}{\gamma}\frac{1}{v_i^2-\frac{1}{4}}+J_{\perp}\right),&{\rm for}~~l=m=1,2,5,\\
-\frac{U_+}{2}-\frac{U_-}{2}+(\frac{J_{\parallel}}{\gamma}-J_{\perp})L\\
+\sum_{i=1}^{N}\left(\frac{J_{\parallel}}{\gamma}\frac{1}{v_i^2-\frac{1}{4}}+J_{\perp}\right),&{\rm for}~~l=m=3,4.
\end{array}\right.
\label{energy}
\end{eqnarray}
\section{Boundary impurity effects }
\label{sec4}

Having diagonalised the Hamiltonian (\ref{Ham}) by means of the algebraic Bethe ansatz, the next step is to derive the thermodynamic Bethe ansatz equations.
\subsection{Derivation of TBA}

For later  convenience in the analysis of the Bethe ansatz equations, we make the change of  variables: $v_i\rightarrow  -\mathrm{i}v_i,~\mu_l\rightarrow -\mathrm{i}\mu_l,~w_k\rightarrow  -\mathrm{i}w_k$ and some rescalings in the boundary parameters $\xi_{\pm}$. The Bethe ansatz equations are now
{\small 
\begin{eqnarray}
&&\zeta(v_i,\beta_{\pm})
\prod_{r=\pm}\prod^{N}_{\stackrel{\scriptstyle l=1}{l\neq i}}\frac{v_i-rv_l-\mathrm{i}}{v_i-rv_l+\mathrm{i}}\prod^{M}_{l=1}\frac{v_i-r\mu_l+\frac{\mathrm{i}}{2}}{v_i-r\mu_l-\frac{\mathrm{i}}{2}}=
\frac{(v_i-\frac{\mathrm{i} }{2})^{2L}}{(v_i+\frac{\mathrm{i}}{2})^{2L}} \label{Beth1},\\
& &\eta (\mu_j,\beta_{\pm})\prod_{r=\pm}\prod^{M}_{\stackrel{\scriptstyle i=1}{i\neq j}}\frac{\mu_j-r\mu_i-\mathrm{i}}{\mu_j-r\mu_i+\mathrm{i}}
\prod^{Q}_{l=1}\frac{\mu_j-rw_l+\frac{\mathrm{i}}{2}}{\mu_j-rw_l-\frac{\mathrm{i}}{2}}\prod^{N}_{i=1}\frac{\mu_j-rv_i+\frac{\mathrm{i}}{2}}{\mu_j-rv_i-\frac{\mathrm{i}}{2}}=1, \label{Beth2}\\
& &\Omega (w_k,\beta_{\pm})\prod_{r=\pm}\prod^{Q}_{\stackrel{\scriptstyle l=1}{l\neq k}}\frac{w_k-rw_l-\mathrm{i}}{w_k-rw_l+\mathrm{i}}\prod^{M}_{l=1}\frac{w_k-r\mu_l+\frac{\mathrm{i}}{2}}{w_k-r\mu_l-\frac{\mathrm{i}}{2}}=1,\label{Beth3}
\end{eqnarray}
}
where
\begin{eqnarray}
& &\zeta(v_{i},\beta_{\pm}) =
\left\{\begin{array}{ll}
\frac{v_i+\mathrm{i}\beta_{\pm}}{v_i-\mathrm{i}\beta_{\pm}},& {\rm for}~~l=1,3,5,~~~m=1,3,5,\\
1,& {\rm for}~~l=2,4,~~~m=2,4,
\end{array}\right.\\
& &
\eta(\mu_{j},\beta_{\pm}) =
\left\{\begin{array}{ll}
1,& {\rm for}~~l=1,2,3,~~~m=1,2,3,\\
\frac{\mu_j+\mathrm{i}\beta_{\pm}}{\mu_j-\mathrm{i}\beta_{\pm}},& {\rm for}~~l=4,~~~m=4,\\
\frac{\mu_j-\mathrm{i}(\beta_{\pm}+\frac{1}{2})}{\mu_j+\mathrm{i}(\beta_{\pm}+\frac{1}{2})},& {\rm for}~~l=5,~~~m=5,
\end{array}\right.\\
& &
\Omega(w_{k},\beta_{\pm}) =
\left\{\begin{array}{ll}
\frac{w_k+\mathrm{i}\beta_{\pm}}{w_k-\mathrm{i}\beta_{\pm}},& {\rm for}~~l=2,5,~~~m=2,5,\\
\frac{w_k-\mathrm{i}(\beta_{\pm}+1)}{w_k+\mathrm{i}(\beta_{\pm}+1)},& {\rm for}~~l=3,~~~m=3,\\
1,& {\rm for}~~l=1,4,~~~m=1,4.
\end{array}\right.
\end{eqnarray}
The shifts in the parameters $\xi_{\pm}$ are given by 
\begin{eqnarray}
\begin{array}{ll}
\beta_{\pm}=\xi_{\pm}-\frac{1}{2},&\,\,{\rm for}\, l=m=1,4,5,\\
\beta_{\pm}=\xi_{\pm}-\frac{3}{2},&\,\,{\rm for}\, l=m=2,\\
\beta_{\pm}=-\xi_{\pm}-\frac{1}{2},&\,\,{\rm for}\, l=m=3.
\end{array}
\end{eqnarray}
Correspondingly, the energy spectrum is given by 
\begin{eqnarray}
E & = &\sum_{i=1}^{N}\left(-\frac{J_{\parallel}}{\gamma}\frac{1}{v_i^2+\frac{1}{4}}+J_{\perp}\right).
\label{energy2}
\end{eqnarray}
Here we have  dropped  some  constants appearing in  Eq. (\ref{energy}), which will be used in deriving one point correlation functions later.

From the above Bethe ansatz equations (\ref{Beth1})-(\ref{Beth3}), it is found that in the cases $l=m=1,3,5$, the
solutions $v_{{\rm l}}=\pm \mathrm{i}\beta_{-}$ and $v_{{\rm r}}=\pm
\mathrm{i}\beta_{+}$ form two boundary bound sates in the charge rapidity
when $\beta_{\pm}$ are negative. Nevertheless, in the case $l=m=5$,
besides the charge boundary bound states,  the boundary
bound states exist also in the spin rapidites, i.e. 
\begin{eqnarray}
\begin{array}{ll}
\mu=\left\{\begin{array}{l}
\pm \mathrm{i}(\beta_-+\frac{1}{2}),\\
 \pm \mathrm{i}(\beta_++\frac{1}{2}),
\end{array}\right.
&
w=\left\{\begin{array}{l}
\pm \mathrm{i}\beta_{-},\\
\pm \mathrm{i}\beta_{+}.\end{array}
\right.
\end{array}\nonumber
\end{eqnarray}
No boundary bound state exists in the remaining cases. We observe  that when
$J_{\perp}>\frac{2J_{\parallel}}{\gamma}(1-\cos k)$, the reference state
becomes the  true ground state, i.e., the ground state is given by a product of
the singlet rung states. The minimal gap can be easily calculated and  is  given by
\begin{equation}
\Delta =J_{\perp}-\frac{2J_{\parallel}}{\gamma}(1-\cos k),
\end{equation}
where $k=\pi
/[1+\frac{1}{4L}(\frac{1}{\beta_+}+\frac{1}{\beta_-})]$. It is obvious
that gap remains almost unchanged in the thermodynamic limit and is
almost the same as  $\Delta
=J_{\perp}-\frac{4J_{\parallel}}{\gamma}$ in the periodic case because $L
>> \frac{1}{\beta _{\pm}}$. In the regime $-\frac{1}{2}<\beta
_{\pm}<-\frac{1}{2}\sqrt{1-\frac{4J_{\parallel}}{\gamma J_{\perp}}}$,
the boundary bound states are stable. Otherwise, in the remaining
regime, they become excited states. In the limit $J_{\perp}\rightarrow
\infty$, all the boundary bound states are excitations.  We shall see
that the boundary bound states radically affect the edge ground state properties.  For $
J_c^-=-\frac{J_{\parallel}}{\gamma}(\frac{\pi}{\sqrt{3}}-\ln
3)<J_{\perp}< \frac{4J_{\parallel}}{\gamma}$, the ground state
consists of three branches of Luttinger liquids associated with the
rapidities $v$,\, $\mu$ and $w$. Here $J_c^-$ is the critical transition point
from the $SU(3)$ phase into the $SU(4)$ phase  in the absence of
a magnetic field. The triplet states can exist in the ground state. This
corresponds to a continuum of massless excitations. 

The thermodynamics of the boundary fields can be derived from the Bethe
ansatz equations (\ref{Beth1})-(\ref{Beth3}). We now 
 focus  on the analysis of the Bethe ansatz equations. As usual, we define  the functions
\begin{eqnarray}
& &e_{n}(x)=\frac{x+\mathrm{i}\frac{n}{2}}{x-\mathrm{i}\frac{n}{2}},~
~~~\theta_{n}(x) = 
 \mathrm{i}\ln e_{n}(x) ,\nonumber\\
& &
a_n(x)=\frac{1}{2\pi}\frac{n}{x^2+\frac{n^2}{4}}\equiv \frac{1}{2\pi}\frac{d}{dx}\theta_{n}(x),
\end{eqnarray}
in terms of which  the  Bethe ansatz equations (\ref{Beth1})-(\ref{Beth3}) become
\begin{eqnarray}
&&\zeta (v_i,\beta_{\pm})(e_1(v_i))^{2L}\prod_{r=\pm}\prod^{M}_{l=1}
e_1(v_i-r\mu_l)=\prod_{r=\pm}\prod^{N}_{\stackrel{\scriptstyle l=1}{l\neq i}}e_2(v_i-rv_l),\label{TBA1}\\
& &
\eta(\mu_j,\beta_{\pm})\prod_{r=\pm}\prod_{l=1}^{N}e_1(\mu_j-rv_l)=\prod_{r=\pm}\prod^{M}_{\stackrel{\scriptstyle l=1}{l\neq j}}e_2(\mu_j-r\mu_l)\prod_{l=1}^{Q}e_{-1}(\mu_j-rw_l),\label{TBA2}\\
& &\Omega(w_k,\beta_{\pm})\prod_{r=\pm}\prod^{M}_{l=1}e_1(w_k-r\mu_l)=\prod_{r=\pm}\prod^{Q}_{\stackrel{\scriptstyle l=1}{l\neq k}}e_2(w_k-rw_l).\label{TBA3}
\end{eqnarray}
In order to study the thermodynamics of the model with boundary impurities we begin by  adopting the
string  hypothesis \cite{TBA}.  If we define $v_{-j}=-v_j$, $\mu_{-l}=-\mu_{l}$ and $w_{-k}=w_{k}$,
 the Bethe ansatz equations
(\ref{Beth1})-(\ref{Beth3}) admit the string solutions
\begin{eqnarray}
v^n_{\alpha_1 j}&=&v^n_{\alpha_1}+\mathrm{i}\frac{1}{2}(n+1-2j),\nonumber\\
\mu^n_{\alpha_2 j}&=&\mu^n_{\alpha_2}+\mathrm{i}\frac{1}{2}(n+1-2j)\nonumber,\\
w^n_{\alpha_3 j}&=&w^n_{\alpha_3}+\mathrm{i}\frac{1}{2}(n+1-2j),\nonumber
\end{eqnarray}
in thermodynamic limit,
where $j=1,\cdots,n$, $\alpha_a =1,\cdots,N_n^{(a)}$ and $v^n_{\alpha_1}$, $\mu^n_{\alpha_2}$ and $w^n_{\alpha_3}$ are  the
positions of the center of the strings. The number
of $n$-strings $N_n^{(a)}$ satisfy the relation $P^{(a)}=\sum_nnN_n^{(a)}$.
By taking the thermodynamic limit, the Bethe ansatz equations become
\begin{eqnarray}
\rho^{(1)h}_n&=&a_n+\frac{1}{2L}\rho^{(1)}_{{\rm b}n}-\sum_{m}A_{nm}*\rho^{(1)}_m+\sum_{m}a_{nm}*\rho^{(2)}_m,\label{Tbethe1}\\ 
\rho^{(2)h}_n&=&\frac{1}{2L}\rho^{(2)}_{{\rm b}n}-\sum_{m}A_{nm}*\rho^{(2)}_m+\sum_{m}a_{nm}*(\rho^{(1)}_m+\rho^{(3)}_m),\label{Tbethe2}\\
\rho^{(3)h}_n&=&\frac{1}{2L}\rho^{(3)}_{{\rm b}n}-\sum_{m}A_{nm}*\rho^{(3)}_m+\sum_{m}a_{nm}*\rho^{(2)}_m,\label{Tbethe3}
\end{eqnarray}
where 
the symbol $*$ denotes the usual convolution. Here $\rho
^{(a)}_n(v), \,a=1,2,3$ are the densities of roots of the three flavors, $\rho
^{(a)h}_n(v),\,a=1,2,3$ are the densities of holes of the three flavors and 
$\rho_{{\rm b}n}^{(i)}, ~i=1,2,3$ are the contributions from boundary fields associated with different rapidities. These boundary phase factors are given by 
\begin{eqnarray}
& & \rho^{(1)}_{{\rm b}n}=
\left\{\begin{array}{ll}
\sum_{\pm}\sum_{l=1}^{n}a_{n+2\beta_{\pm}+1-2l}(\lambda )+a_{n2}(\lambda),& {\rm for}~~l=1,3,5,~~m=1,3,5,\\
a_{n2}(\lambda),& {\rm for}~~l=2,4,~~m=2,4,
\end{array}\right.\\
& &
 \rho^{(2)}_{{\rm b}n}=
\left\{\begin{array}{ll}
a_{n2}(\lambda)-a_{n1}(\lambda),& {\rm for}~l=1,2,3,~m=1,2,3,\\
\sum_{\pm}\sum_{l=1}^{n}a_{n+2\beta_{\pm}+1-2l}(\lambda ) +a_{n2}(\lambda)-a_{n1}(\lambda),& {\rm for}~l=4,~m=4,\\
-\sum_{\pm}\sum_{l=1}^{n}a_{n+2\beta_{\pm}+2-2l}(\lambda )\\
+a_{n2}(\lambda)-a_{n1}(\lambda),& {\rm for} ~~l=5,~~~m=5,
\end{array}\right.\\
& &
\rho^{(3)}_{{\rm b}n} =
\left\{\begin{array}{ll}
\sum_{\pm}\sum_{l=1}^{n}a_{n+2\beta_{\pm}+1-2l}(\lambda )+a_{n2}(\lambda),& {\rm for}~~l=2,5,~~m=2,5,\\
-\sum_{\pm}\sum_{l=1}^{n}a_{n+2\beta_{\pm}+3-2l}(\lambda )+a_{n2}(\lambda),& {\rm for}~~l=3,~~m=3,\\
a_{n2}(\lambda),& {\rm for}~~l=1,4,~~m=1,4.
\end{array}\right.
\end{eqnarray}
In addition
\begin{eqnarray}
A_{nm}(\lambda)&=& \delta(\lambda)\delta_{nm}+(1-\delta_{nm})a_{|n-m|}(\lambda)
+a_{n+m}(\lambda)\nonumber\\
& &
\qquad
+2\sum^{{\rm
Min}(n,m)-1}_{l=1}a_{|n-m|+2l}(\lambda),\nonumber\\
a_{nm}(\lambda) &=& \sum^{{\rm Min}(n,m)}_{l=1}a_{n+m+1-2l}(\lambda).\nonumber
\end{eqnarray}
We emphasize that the boundary potentials enter in the expression for
the ground state energy implicitly via $\rho^{(a)}_{{\rm b}}(v)$ in the above equations, with contributions to the densities of the roots at the
order of $1/L$. In order to find the equilibrium state of the system at fixed temperature $T$ and external magnetic field $H$ ($\geq 0$), we  minimize  the free energy $F=E-TS-HS^z$ with respect
to the densities to obtain the TBA in the form
{\small
\begin{flushleft}
\begin{equation}
\left(\begin{array}{c}
\ln(1+\eta _n^{(1)})\\
\ln(1+\eta _n^{(2)})\\
\ln(1+\eta _n^{(3)})
\end{array}\right)
=\frac{G_n}{T}+\left(\begin{array}{ccc}
\sum_m A_{nm}&-\sum_m a_{nm}&0\\
-\sum_m a_{nm}&\sum_m A_{nm}&-\sum_m a_{nm}\\
0&-\sum_m a_{nm}&\sum _{m}A_{nm}
\end{array}\right)*\left(\begin{array}{l}
\ln(1+\frac{1}{\eta _m^{(1)}})\\
\ln(1+\frac{1}{\eta _m^{(2)}})\\
\ln(1+\frac{1}{\eta _m^{(3)}})
\end{array}\right).\label{TBA}
\end{equation}
\end{flushleft}
}
The driving matrix  $G_n$ depends on the choice of
the reference state. Explicitly, for $J_{\perp}<0$, $G={\rm column}(-\frac{J_{\parallel}}{\gamma} 2\pi a_n+nH,nH,-n(J_{\perp}+H))$, giving the free energy
\begin{eqnarray}
\frac{F(T,H)}{2L}& =& -H-T\int_{-\infty}^{\infty}\sum_{n=1}^{\infty}a_n(\lambda)\ln(1+e^{-\frac{\epsilon^{(1)}_n(\lambda)}{T}})d\lambda\nonumber\\
& &-\frac{T}{2L}\sum_{a=1}^{3}\int_{-\infty}^{\infty}\sum_{n=1}^{\infty}\rho_{{\rm b}n}^{(a)}
\ln(1+e^{-\frac{\epsilon^{(a)}_n(\lambda)}{T}})d\lambda. \label{FE1}
\end{eqnarray}
On the other hand, for $J_{\perp}>0$, $G={\rm colum}(-\frac{J_{\parallel}}{\gamma}2\pi a_n+n(J_{\perp}-H), nH,nH)$ and 
the free  energy is  given by 
\begin{eqnarray}
\frac{F(T,H)}{2L}& = &-T\int_{-\infty}^{\infty}\sum_{n=1}^{\infty}a_n(\lambda)\ln(1+e^{-\frac{\epsilon^{(1)}_n(\lambda)}{T}})d\lambda\nonumber\\
& &-\frac{T}{2L}\sum_{a=1}^{3}\int_{-\infty}^{\infty}\sum_{n=1}^{\infty}\rho_{{\rm b}n}^{(a)}
\ln(1+e^{-\frac{\epsilon^{(a)}_n(\lambda)}{T}})d\lambda. \label{FE2}
\end{eqnarray}
Here
$\eta ^{(l)}_n(\lambda)=\rho
^{(l)h}(\lambda)/\rho ^{(l)}(\lambda)\equiv
{\rm exp}(\epsilon^{(l)}_n(\lambda)/T),\, l=1,2,3$, with the  dressed energy
$\epsilon^{(l)}_n$ playing  the role of an excitation energy measured
from the Fermi level. 

Using the relations,
\begin{eqnarray}
(a_0+a_2)*\ln \eta_n^{(a)} &=&a_1*\left[\ln (1+\eta^{(a)}_{n+1})+\ln (1+\eta^{(a)}_{n-1})\right]\nonumber\\
& &-\ln (1+\frac{1}{\eta^{(a-1)}_{n}})-\ln (1+\frac{1}{\eta^{(a+1)}_{n}}),
\end{eqnarray}
another form of the TBA is given by
\begin{eqnarray}
\epsilon^{(a)}_1&=&g_1^{(a)}+Ta_2*\ln(1+e^{-\frac{\epsilon^{(a)}_1}{T}})+
T(a_0+a_2)\sum_{m=1}^{\infty}a_m*\ln(1+e^{-\frac{\epsilon^{(a)}_{m+1}}{T}})\nonumber\\
& &-T\sum_{m=1}^{\infty}a_m*\left(\ln(1+e^{-\frac{\epsilon^{(a-1)}_m}{T}})+\ln(1+e^{-\frac{\epsilon^{(a+1)}_m}{T}})\right),\label{TBAe1}\\
\epsilon^{(a)}_n&=&g_n^{(a)}+Ta_1*\ln(1+e^{\frac{\epsilon^{(a)}_{n-1}}{T}})\nonumber\\
& &+
Ta_2*\ln(1+e^{-\frac{\epsilon^{(a)}_n}{T}})+T(a_0+a_2)\sum_{m\geq n}^{\infty}a_{m-n}*\ln(1+e^{-\frac{\epsilon^{(a)}_m}{T}})\nonumber\\
& &-T\sum_{m\geq n}^{\infty}a_{m-n+1}*\left(\ln(1+e^{-\frac{\epsilon^{(a-1)}_m}{T}})+\ln(1+e^{-\frac{\epsilon^{(a+1)}_m}{T}})\right),\label{TBAe2}
\end{eqnarray}
for $n\geq 2$.
In the above $a=1,2,3$ and $\epsilon^{(0)}_n(\lambda)=\epsilon^{(4)}_n(\lambda)=0$ is assumed. The driving terms are given explicitly by
\begin{eqnarray}
\begin{array}{lll}
\begin{array}{l}
g_1^{(1)}=-\frac{J_{\parallel}}{\gamma} 2\pi a_1+H,\\
g_1^{(2)}=H,\\
g_1^{(3)}=-(J_{\perp}+H),
\end{array} &
\begin{array}{l}
g_n^{(1)}=H,\\
g_n^{(2)}=H,\\
g_n^{(3)}=-(J_{\perp}+H),
\end{array} 
&~~~~~{\rm for}~~J_{\perp}<0,
\end{array}\\
\begin{array}{lll}
\begin{array}{l}
g_1^{(1)}=-\frac{J_{\parallel}}{\gamma} 2\pi a_1+J_{\perp}-H,\\
g_1^{(2)}=H,\\
g_1^{(3)}=H,
\end{array} &
\begin{array}{l}
g_n^{(1)}=J_{\perp}-H,\\
g_n^{(2)}=H,\\
g_n^{(3)}=H.
\end{array}
&
~~~~~{\rm for}~~J_{\perp}\geq 0.
\end{array}
\end{eqnarray}
\subsection{Boundary bound states and impurity effects}
In the low temperature limit, the states with positive dressed energy are empty. The zeros of the dressed energies define the fermi energies. We decompose $\epsilon^{(a)}_n$ into its positive and negative parts,
$\epsilon^{(a)}_n=\epsilon^{(a)+}_n+\epsilon^{(a)-}_n$. An analysis of equations (\ref{TBAe1}) and (\ref{TBAe2}) in the limit $T\rightarrow 0$ reveals that for the ground state, the roots are all real corresponding  to $n=1$. All dressed energies $\epsilon^{(a)+}_n$ with $n\geq 2$ correspond to excitations. Thus the TBA for the ground state is, for $J_{\perp} <0$,
\begin{eqnarray}
\epsilon^{(1)}&=&g_1^{(1)}-a_2*\epsilon^{(1)-}+a_1*\epsilon^{(2)-},\nonumber\\
\epsilon^{(2)}&=&H-a_2*\epsilon^{(2)-}+a_1*\left[\epsilon^{(1)-}+\epsilon^{(3)-}\right], \nonumber\\
\epsilon^{(3)}&=&-H-J_{\perp}-a_2*\epsilon^{(3)-}+a_1*\epsilon^{(2)-},\label{TBAg1}
\end{eqnarray}
and for $J_{\perp} \geq 0$,
\begin{eqnarray}
\epsilon^{(1)}&=&g_1^{(1)}-a_2*\epsilon^{(1)-}+a_1*\epsilon^{(2)-},\nonumber\\
\epsilon^{(2)}&=&H-a_2*\epsilon^{(2)-}+a_1*\left[\epsilon^{(1)-}+\epsilon^{(3)-}\right], \nonumber\\
\epsilon^{(3)}&=&H-a_2*\epsilon^{(3)-}+a_1*\epsilon^{(2)-}.\label{TBAg2}
\end{eqnarray}
In this case,
the free energy is given by 
\begin{equation}
\frac{F(0,H)}{2L}=\left\{
\begin{array}{ll}
-H+\int_{-\infty}^{\infty}a_1(\lambda)\epsilon_1^{(1)-}(\lambda )d\lambda+\frac{1}{2L}f_{{\rm b}},&~~{{\rm for}}~~J_{\perp} < 0,\\
\int_{-\infty}^{\infty}a_1(\lambda)\epsilon_1^{(1)-}(\lambda )d\lambda+\frac{1}{2L}f_{{\rm b}},&~~{{\rm for}}~~J_{\perp}\geq 0,
\end{array}\right. \label{FEg}
\end{equation}
where 
\begin{equation}
f_{{\rm b}}=\sum_{a=1}^3\int_{-Q_a}^{Q_a}\rho_{{\rm b}1}^{(a)}(\lambda)\epsilon_1^{(a)-}(\lambda)+\theta(\beta_{\pm}+\beta_c)E_{{\rm bs}},
\end{equation}
and $\theta(x)$ denotes  a step-like  function. Define $\beta_c=-\frac{1}{2}\sqrt{1-\frac{4J_{\parallel}}{\gamma J_{\perp}}}$, then  in the interval $ -\frac{1}{2}<\beta< \beta_c$,  $\theta(\beta )=1$, else $\theta(\beta )=0$. In the above  $E_{{\rm bs}}$ denotes the boundary bound state energy,  given by 
\begin{equation}
E_{{\rm bs}}=\left\{
\begin{array}{ll}
\sum_{\pm}\left(-\frac{J_{\parallel}}{\gamma}\frac{1}{-\beta_{\pm}^2+\frac{1}{4}}+J_{\perp}\right),&~~{\rm for}~J_{\perp}\geq 0\\
\sum_{\pm}\left(-\frac{J_{\parallel}}{\gamma}\frac{1}{-\beta_{\pm}^2+\frac{1}{4}}\right),&~~{\rm for}~J_{\perp}<0.\end{array}\right.
\end{equation}
It is worth noticing that if $\beta_{\pm} <\beta_c$ we should take the boundary bound states into account in the boundary contributions $\rho_{{\rm b}1}^{(a)}$ for the cases $l=m=1,3,5$. 
The TBA (\ref{TBAg1}) and (\ref{TBAg2}) provide a clear  physical picture of the ground-state and in turn the thermodynamic properties, such as the 
free energy, the magnetization, the susceptibility, etc.  The boundary impurities coupled to the three rapidities affect the low temperature physics at the edges in various different  ways, which we now explore. 

From the TBA (\ref{TBAg2}), we notice that if
$J_{\perp}>J_c^+=\frac{4J_{\parallel}}{\gamma}$ the triplet
excitations are massive with  energy gap
$\Delta=J_{\perp}-\frac{4J_{\parallel}}{\gamma}$. The rescaling
$\gamma =4$ was fixed \cite{BGFZ} for strong coupling compounds, e.g.
Cu$_{2}$(C$_5$H$_{12}$N$_2$)$_2$Cl$_4$ \cite{exp3}, (C$_5$H$_{12}$N)$_2$CuBr$_{4}$ \cite{exp4}, etc.
Here $J^+_c$ is the critical point at which the quantum phase
transition from the three branches of Luttinger liquid  to the
dimerized $U(1)$ phase occurs.  If $J_{\perp} >J_c^+$, we can show
that in the presence of a strong magnetic field two of the triplet
states ($|3\rangle $ and $|4\rangle $ in (\ref{basis})) in the bulk part
will never be involved in the ground state. However, at the boundaries
this  is not always true due to the presence of the boundary impurities.  In
a strong magnetic field the ground-state may be considered as a
condensate of $SU(2)$ hard-core bosons. The gap is reduced by the
magnetic field $H$,
i.e. $\Delta=J_{\perp}-\frac{4J_{\parallel}}{\gamma}-H$. Thus the
first critical field occurs at the point  $H_{c1}$ where the gap is closed, i.e.  $g\mu_BH_{c1}= J_{\perp}-\frac{4J_{\parallel}}{\gamma} $. The quantum phase transition
from a gapped phase to gapless Luttinger phase occurs. By continuing to increase the magnetic
field $H$ over  $H_{c1}$, the triplet state $|2\rangle$ becomes
involved in the ground state with a finite susceptibility,  also
affected by the boundary impurities in the low concentration regime. 
If the magnetic field is greater than the rung
coupling, i.e. $h>J_{\perp}$, the triplet component $|2\rangle$ becomes the lowest
level. Therefore, it is reasonable to choose the basis order as
$(|2\rangle,|1\rangle,|3\rangle,|4\rangle)^{{\rm T}}$. Subsequently
the driving terms are given by $g^{(1)}=-2\pi
J_{\parallel}a_1-J_{\perp}+H$, $g^{(2)}=J_{\perp}$ and
$g^{(3)}=H$. A second critical field $H_{c2}$ ($H_{c2}>H_{c1}$) can be determined by the magnetization arriving at its saturation value
$S^z=1$. Then the reference state becomes the true physical state and  the
critical field $H_{c2}$ is given by
\begin{equation}
H_{c2}=J_{\perp}+\frac{4J_{\parallel}}{\gamma}.\label{Hc2}
\end{equation}
In this case, all the boundary impurities are gapfull with the ferromagnetic gap
$\Delta =\mu_B g(H-H_{c2})$.

Let us now discuss the boundary impurity effects in the vicinity of the critical point $H_{c1}$. After a lengthy calculation, similar to that employed in \cite{BGFZ} for the periodic case, we find the free energy in the presence of a strong  magnetic field $H$,
\begin{equation}
\frac{F(0,H)}{2L}\approx -\frac{4Q(J_c^+-J_{{\rm eff}})}{\pi}\left(1-\frac{2Q}{\pi}\right)+\frac{1}{2L}f_{{\rm b}},
\end{equation}
where $Q$ is the fermi point given by 
$Q\approx \sqrt{\frac{J_c^+-J_{{\rm eff}}}{4J_c^+-5(H-H_{c1})}}$ and $ f_{{\rm b}}$ is the  surface free energy from the  boundary impurities in the vicinity of $H_{c1}$.  Explicitly,\\
for $|\beta_{\pm}| \geq \frac{1}{2}$, or say $ 0<U_{\pm}\leq \frac{J_{\parallel}}{\gamma }$ or $ U_{\pm}<0$, it is given by 
\begin{equation}
f_{{\rm b}}\approx
\left\{\begin{array}{ll}
-\frac{2Q(J_c^+-J_{{\rm eff}})}{\pi}(1+\frac{1}{\beta_+}+\frac{1}{\beta_-}),&~~{{\rm for }}~~l=m=1,3,5, \\
-\frac{2Q(J_c^+-J_{{\rm eff}})}{\pi},&~~{{\rm for }}~~l=m=2,4.\end{array}\right.\nonumber
\end{equation}
For $-\frac{1}{2}<\beta_{\pm} << \beta_c$, or  say $ U_{\pm}>>U_{{\rm bs}}= 2J_{\parallel}/\gamma (1-\sqrt{1-\frac{4J_{\parallel}}{\gamma J_{\perp}}})$, we have 
\begin{equation}
f_{{\rm b}}\approx
\left\{\begin{array}{ll}
-\sum_{\pm}\frac{2Q(J_c^+-J_{{\rm eff}})}{\pi}\left(\frac{3}{2}+\frac{1}{\beta_{\pm}}+\frac{1}{\beta_{\pm}+1}-\frac{1}{\beta_{\pm}-1}\right)\\
+\sum_{\pm}\left(-\frac{J_{\parallel}}{\gamma}\frac{1}{-\beta_{\pm}^2+\frac{1}{4}}+J_{\perp}\right),&~{{\rm for}}~l=m=1,3,5,\nonumber\\
-\frac{2Q(J_c^+-J_{{\rm eff}})}{\pi},&~{{\rm for }}~l=m=2,4.
\end{array}\right.\nonumber
\end{equation}
While for $|\beta_{\pm}|$ very small, or say $\frac{J_{\parallel}}{\gamma}<U_{\pm}< U_{{\rm bs}}$, it is given by
\begin{equation}
f_{{\rm b}}\approx
\left\{\begin{array}{ll}
-\frac{2(J_c^+-J_{{\rm eff}})Q}{\pi}+\sum_{\pm}f(\beta_{\pm})&~{{\rm for }}~l=m=1,3,5,\nonumber\\
-\frac{2Q(J_c^+-J_{{\rm eff}})}{\pi}&~~{{\rm for}}~~l=m=2,4,\end{array}\right. \nonumber
\end{equation}
where 
\begin{eqnarray}
f(\beta_{\pm})&=&-\frac{8J_{\parallel}}{\pi \gamma}\frac{1}{1-4\beta_{\pm}^2}\left(\arctan{\frac{Q}{\beta_{\pm}}}-4Q\beta_{\pm}\right)+\frac{2J_{{\rm eff}}}{\pi}\arctan{\frac{Q}{\beta_{\pm}}}\nonumber\\
& &+\frac{4Q(J_c^+-J_{{\rm eff}})}{\pi (\pi +2Q)}\frac{1}{1-\beta_{\pm}^2}\left(\arctan{\frac{Q}{\beta_{\pm}}}-\beta_{\pm}Q\right).
\end{eqnarray}
In the above $J_{{\rm eff}}=J_{\perp}-H$ and the parameters $\beta _{\pm}$ are related to the boundary impurity coupling $U_{\pm}$  by
\begin{equation}
\frac{1}{\beta_{\pm}}=\left\{
\begin{array}{ll}
\frac{2U_{\pm}}{\frac{2J_{\parallel}}{\gamma}-U_{\pm}},&\,\,{\rm for}\, l=m=1,4,5,\\
\frac{2U_{\pm}}{\frac{2J_{\parallel}}{\gamma}-3U_{\pm}},&\,\,{\rm for}\, l=m=2,\\
-\frac{2U_{\pm}}{\frac{2J_{\parallel}}{\gamma}+U_{\pm}},&\,\,{\rm for}\, l=m=3.
\end{array}
\right.\label{relation}
\end{equation}
The magnetic susceptibility follows from $\chi \approx -\frac{d^2}{dH^2}\frac{F(0,H)}{2L}$. Here, to
illustrate the boundary effects, we will focus on the discussion of the
strong coupling compounds $J_{\perp} >> J_{\parallel}$ with the
boundary impurities in the case $l=m=1$. Other regimes can be handled
in a similar way. It is very clear that the stable boundary bound
states are exhibited  only in the strong ferromagnetic boundary coupling $
U_{\pm}> U_{{\rm bs}}$. In  Eq.(\ref{relation}), we emphasize that  the
mathematical singular points do not exist, or alternatively   $\beta _{\pm}=0$  does not mean that  the rhs of Eq. (\ref{relation})  has  singular points. For instance, if $U_{\pm}=2J_{\parallel}/\gamma$, the boundary parameters $\xi_{\pm} =1/2$. Thus the phase factors in  the Bethe ansatz equations (\ref{Beth1}), (\ref{Beth2}) and (\ref{Beth3}) are equal to $1$ for the case $l=m=1$. In such a case, the model exihibits special symmetry (the quantum algebra $SU_q(4)$ invariant Bethe ansatz equations) which leads to a different expression for  the boundary free energy than the above ones. For antiferromagnetic
boundary coupling $U_{\pm}<0$, the susceptibility is given by
\begin{equation}
\chi \approx \frac{3}{\pi\sqrt{4J_c^+(H-H_{c1})}}\left(1+\frac{1}{4L}\sum_{\pm}(\frac{1}{2}+\frac{1}{\beta_{\pm}})\right),
\end{equation}
while for the strong ferromagnetic coupling $U_{\pm}>> U_{{\rm bs}}$, with $U_{\pm} >0$,
\begin{equation}
\chi \approx \frac{3}{\pi\sqrt{4J_c^+(H-H_{c1})}}\left(1+\frac{1}{4L}\sum_{\pm}(\frac{3}{2}+\frac{1}{\beta_{\pm}}+\frac{1}{\beta_{\pm}+1}-\frac{1}{\beta_{\pm}-1})\right).
\end{equation}
Notice, in both cases, that  the susceptibility
diverges with the square root of the field in the bulk and in the
boundaries. In addition,  the susceptibility at
the boundaries is enhanced or decreased by different impurity
dopings. This behaviour is illustrated in figure 2. 
From the Bethe ansatz equations, we can also calculate  the magnetization in the vicinity of $H_{c1}$,
\begin{equation}
\frac{S^z}{2L}=\int_{-Q}^{Q}\rho_1^{(1)}(\lambda )d\lambda=
\frac{4Q}{\pi}(1-\frac{2Q}{\pi})+\frac{1}{2L}\sum _{\pm}S^z_{{\rm b}}.
\end{equation} 
For antiferromagnetic boundary coupling $U_{\pm}<0$ this expression reduces to 
\begin{equation}
S^z_{{\rm b}}\approx \sum_{\pm}\frac{2Q}{\pi}(1-\frac{2Q}{\pi})(\frac{1}{2}+\frac{1}{\beta_{\pm}}),\label{SZ1}
\end{equation}
while for strong  ferromagnetic boundary coupling $U_{\pm}>>U_{{\rm bs}}$ with $U_{\pm}>0$,   
\begin{equation}
S^z_{{\rm b}}\approx \sum_{\pm}\frac{2Q}{\pi}(1-\frac{2Q}{\pi})(\frac{3}{2}+\frac{1}{\beta_{\pm}}+\frac{1}{\beta_{\pm}+1}-\frac{1}{\beta_{\pm}-1}). \label{SZ2}
\end{equation}
\begin{center}
\begin{figure}
\epsfig{file=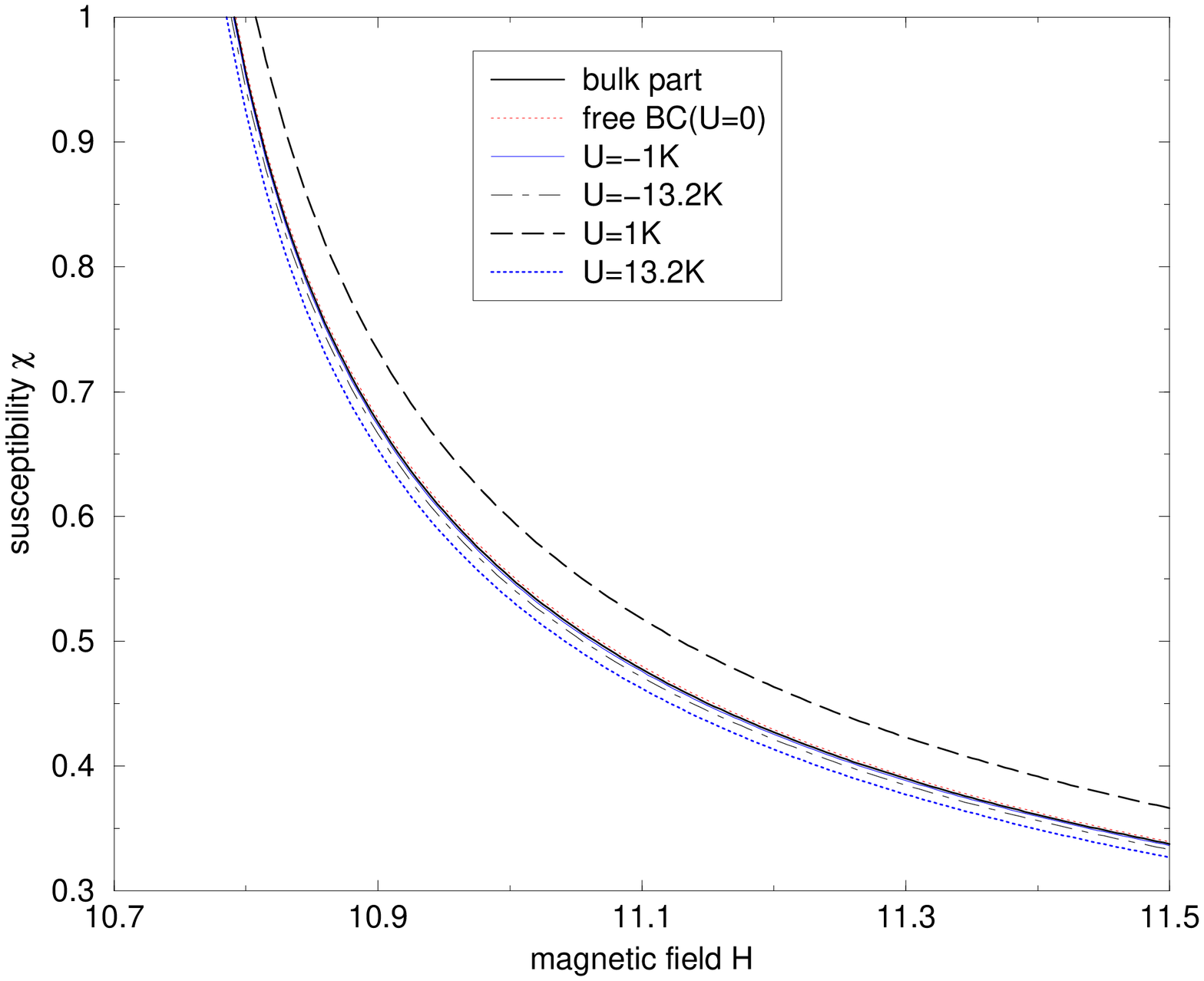,width=7cm,height=12cm,angle=-90}
\caption{The susceptibility versus  magnetic $H$ for different  impurity coupling $U_{\pm}=U$ ($U=0$ corresponds to free boundary conditions). Here we consider the strong coupling compound Cu$_{2}$(C$_5$H$_{12}$N$_2$)$_2$Cl$_4$ \cite{exp3}
with $J_{\perp}=13.2K$, $J_{\parallel}=2.5K$ and $\gamma=4$ with an impurity concentration $2$ percent in a ladder with length $L=50$. }
\end{figure}
\end{center}
\begin{center}
\begin{figure}
\epsfig{file=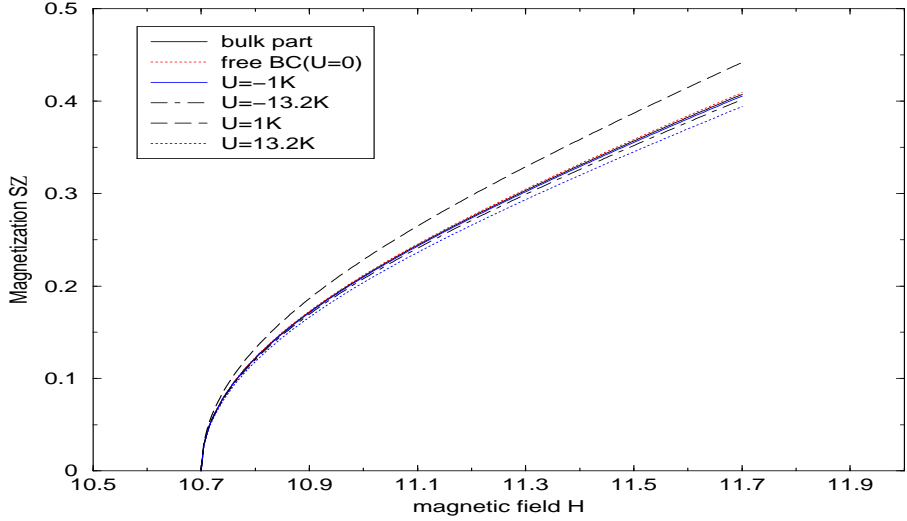,width=7cm,height=12cm,angle=-90}
\caption{The magnetization  versus  magnetic field $H$ for different  impurity coupling $U_{\pm}=U$ ($U=0$ corresponds to free boundary conditions). As in figure 2, we consider the strong coupling compound  Cu$_{2}$(C$_5$H$_{12}$N$_2$)$_2$Cl$_4$ \cite{exp3}
with $J_{\perp}=13.2K$, $J_{\parallel}=2.5K$ and $\gamma=4$ with an impurity concentration $2$ percent in a ladder with length $L=50$. }
\end{figure}
\end{center}
A plot of the magnetization $S^z$ against the magnetic field for different boundary impurities $U_{\pm}$ is given  in figure 3. By analyzing both figures we can observe the  competition between the boundary impurities and the magnetic field in the thermodynamic properties.  In particular, 
 we find an 
enhancement of the susceptibility  in the weak anti- and ferromagnetic  regimes (we consider the sizes $2L=100$, the impurity
concentration $2$ percent ).  The susceptibility and the magnetization are
lifted slightly in the weak antiferromagnetic boundary regime in the case of  open boundary conditions, whereas they contribute negatively to the bulk when $U_{\pm}$ becomes
more and more negative.  This is reasonable, since  negative $U_{\pm}$ energetically favours the
singlet state (recall  the boundary terms in the Hamiltonian (\ref{Ham})),
even if the magnetic field is very strong, such that the spin-$1$ component of
the triplet is  involved in the ground state. The point is that  a very 
negative $U_{\pm}$ can overcome the spin-$1$ component of the triplet
and dominate the edge rung state. In this circumstance, the edge state is a pure singlet state and the edge magnetization (\ref{SZ1}) is zero due to the fact that $U_{\pm}$ effectively decreases the edge magnetic field $H$ to $H_{c1}$ such that the fermi boundary $Q=0$.
This results in negative
susceptibility and magnetization contributions to the bulk. In
contrast to this case, the ferromagnetic impurities lift 
the susceptibility and the magnetization in the weak coupling regime
$U_{\pm}< \frac{J_{\parallel}}{\gamma}$. When $U_{\pm}$ becomes
larger, the triplet edge state is energetically favoured so
that the boundary coupling can overcome the magnetic field to bring
the three components of the triplet into the edge states.  Therefore it causes
a negative contribution to the bulk susceptibility and  magnetization.  This situation is
different from the case of the bulk impurities, where the
susceptibility is increased by the impurity coupling due to the
forward-scattering.  This fact can be seen clearly from the one point
correlation function of the ground state at the edges, for 
antiferromagnetic boundary coupling and weak ferromagnetic boundary coupling, i.e. $U_{\pm}\leq \frac{J_{\parallel}}{\gamma}$,
\begin{eqnarray}
\langle \vec{S }_a.\vec{T }_a\rangle =-\frac{3}{4}+\frac{d}{dU_{\pm}}f_{{\rm b}}
=-\frac{3}{4}+\frac{2Q(H-H_{c1})}{\pi}\frac{4J_{\parallel}/\gamma}{(2J_{\parallel}/\gamma-U_{\pm})^2},\label{CA}
\end{eqnarray}
and for ferromagnetic impurities in the strong coupling regime $U_{\pm}>>P_{{\rm bs}}$,
\begin{eqnarray}
\langle \vec{S }_a.\vec{T }_a\rangle &=&\frac{1}{4}+\frac{2Q(H-H_{c1})}{\pi}\left[\frac{4J_{\parallel}/\gamma}{(2J_{\parallel}/\gamma-U_{\pm})^2}\right. \nonumber\\
& & \left. -\frac{4J_{\parallel}/\gamma}{(2J_{\parallel}/\gamma-3U_{\pm})^2}+\frac{4J_{\parallel}/\gamma}{(2J_{\parallel}/\gamma+U_{\pm})^2}\right]-\frac{1}{(1-\frac{\gamma U_{\pm}}{J_{\parallel}})^2}.\label{CF}
\end{eqnarray}
In the above $a=1,L$.  The boundary one point correlation functions are given by \begin{equation}
\langle \vec{S }_a.\vec{T }_a\rangle =-\frac{3}{4}\langle N_{{\rm S}}\rangle +\frac{1}{4}\langle N_{{\rm T}}\rangle.
\end{equation}
Here $N_{{\rm S}}$ and $N_{{\rm T}}$ are the probabilities of the singlet
and triplet state respectively. This is because the eigenvalue of the one point correlation function $\langle \vec{S }_a.\vec{T }_a\rangle $ acting on the singlet (triplet)  state is $-\frac{3}{4}$ ($\frac{1}{4}$).  We have plotted  
the correlation function for antiferromagnetic boundary
coupling in figure 4. 
\begin{center}
\begin{figure}
\epsfig{file=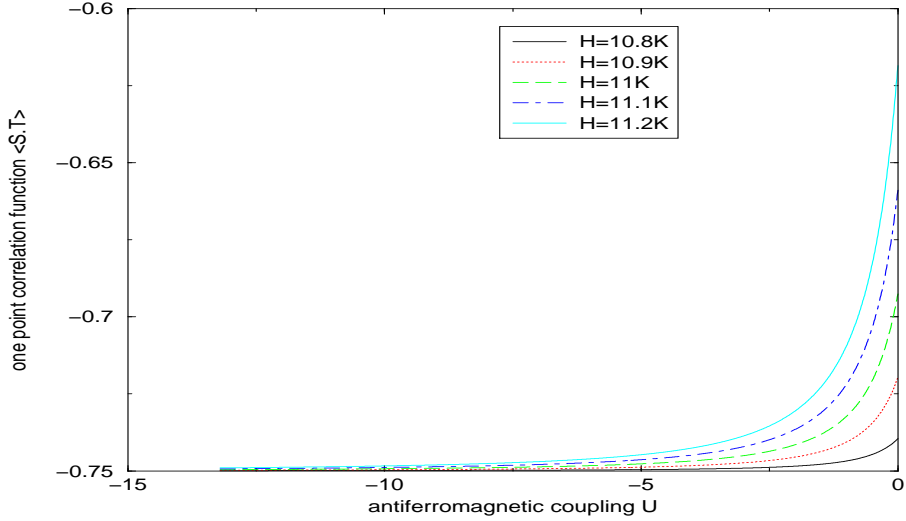,width=7cm,height=12cm,angle=-90}
\caption{One point correlation function (\ref{CA}) versus antiferromagnetic  boundary coupling  $U$ as a function of magnetic field. The curve is lifted by the magnetic field, however it is  decreased by the boundary impurities which favour the singlet state. Here we consider   the strong coupling compound Cu$_{2}$(C$_5$H$_{12}$N$_2$)$_2$Cl$_4$ \cite{exp3} with $J_{\perp}=13.2K$, $J_{\parallel}=2.5K$ and $\gamma=4$ and  $U_{\pm}=U$. The case $U=0$ corresponds to the free boundary effect. }
\end{figure}
\end{center}
\begin{center}
\begin{figure}

\begin{tabular}{cc}
   &       \\
(a) & (b)   \\
\epsfig{file=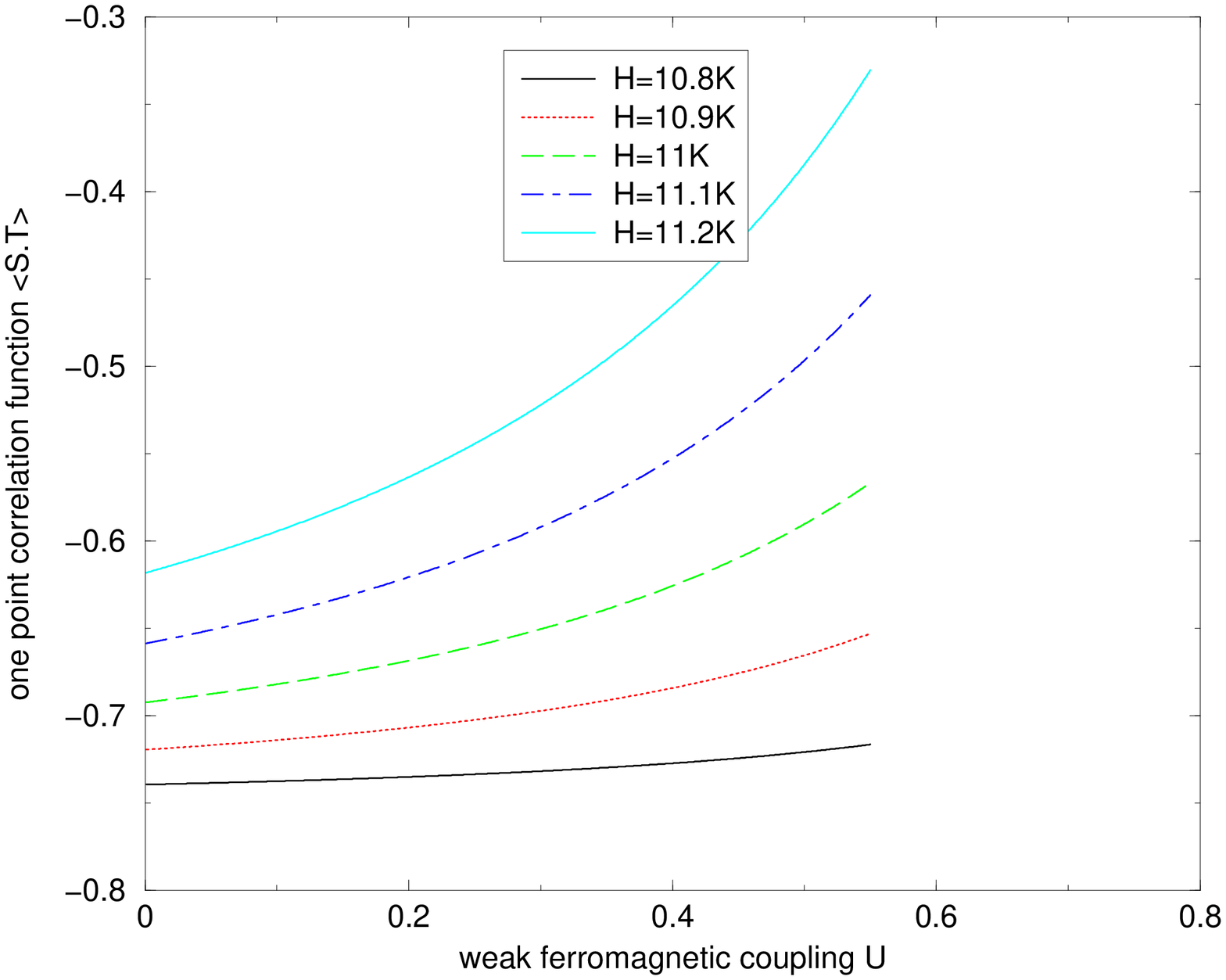,width=5cm,height=6cm,angle=-90}&
\epsfig{file=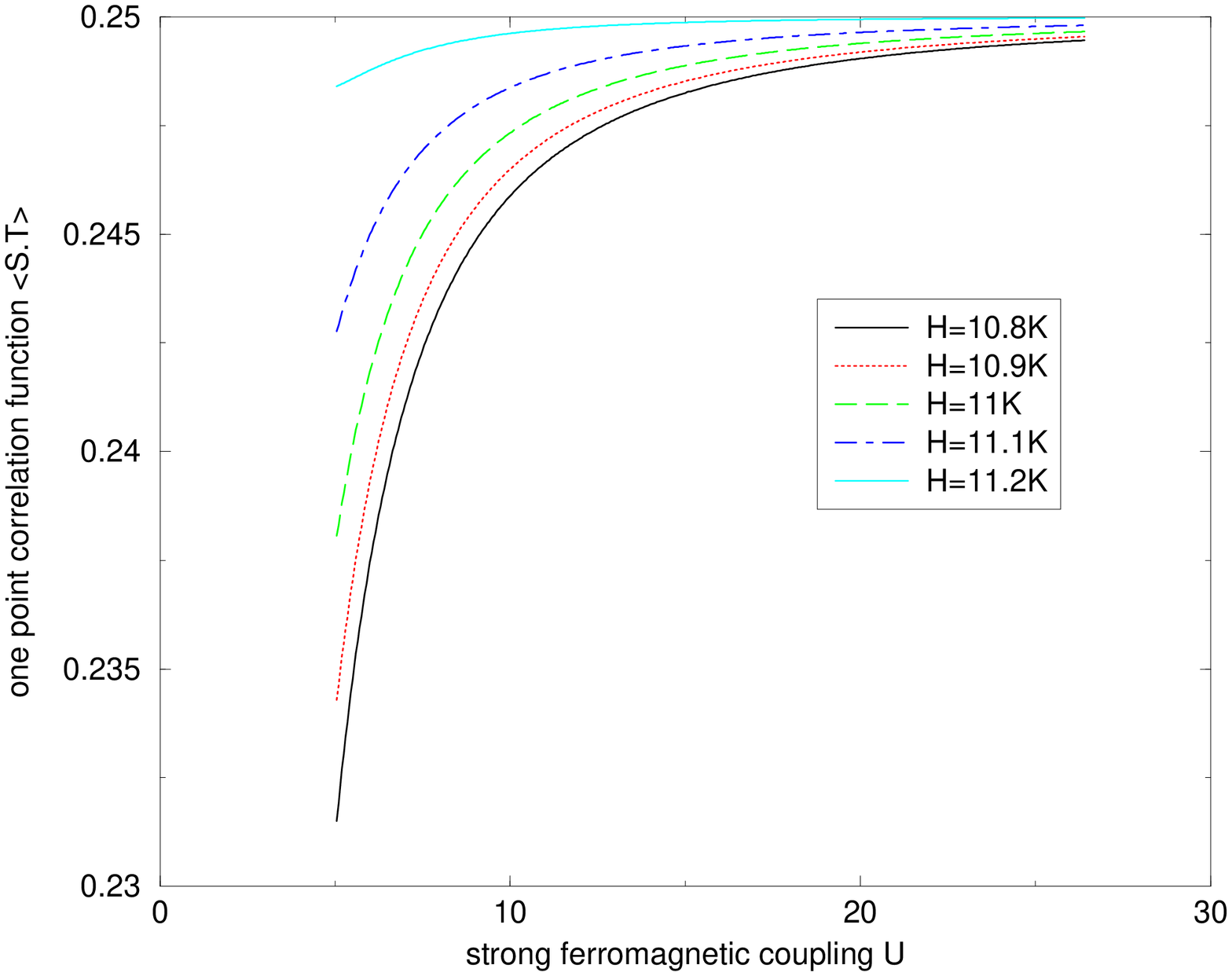,width=5cm,height=6cm,angle=-90}
\end{tabular}
\caption{One point correlation function  vs ferromegnetic boundary coupling $U$:(a) The function (\ref{CA}) is lifted by the magnetic field and weak magnetic impurity coupling $U$. (b) The function (\ref{CF}) tends to $\frac{1}{4}$ as the boundary impurity coupling becomes larger. Here we again consider the strong coupling compound Cu$_{2}$(C$_5$H$_{12}$N$_2$)$_2$Cl$_4$ \cite{exp3} with $J_{\perp}=13.2K$, $J_{\parallel}=2.5K$ and $\gamma=4$ and $U_{\pm}=U$. The case $U=0$ corresponds to free boundaries. }
\end{figure}
\end{center}
We  see that the magnetic field always lifts the spin-$1$ triplet
component. However, in the case of  antiferromagnetic
boundary impurities and open boundaries the singlet state is favoured
as long as $U_{\pm}$ becomes more negative, the triplet moves 
out of the edge state and the one point correlation function tends to
$-\frac{3}{4}$. On the other hand, for ferromagnetic coupling
impurities (see figure 5), the correlation function
increases due to the ferromagnetic properties and the magnetic
field in the weak coupling regime
$U_{\pm}<\frac{J_{\parallel}}{\gamma}$. However, if $U_{\pm}$ becomes larger, the three
components of the triplet get involved in the edge state, such that
the correlation function tends to $\frac{1}{4}$ for strong
ferromagneitc impurity coupling. This result indicates that the edge
state can be a pure singlet state in the  strong antiferromagnetic
boundary coupling regime whereas it turns out to be a pure triplet
state in the strong ferromagnetic boundary coupling regime.  This reveals
the role of anti- and ferromagnetic impurities.

On the other hand, the boundary impurities coupled to the spin degrees of freedom, namely, $\rho_{{\rm b}}^{(2)}$ and $\rho_{{\rm b}}^{(3)}$ will also affect the ground state properties non-trivially. From the free energy (\ref{FEg}),   these impurity densities will contribute to the low energy. Considering  the case $J_{\perp}<0$, in the absence of the magnetic field, the
triplet is completely degenerate while the fermi surface of the singlet
is lifted as $J_{\perp}$ becomes more negative.  Certainly,
if $J_{\perp} <J_c^{-}=-\frac{J_{\parallel}}{\gamma}(\frac{\pi}{\sqrt{3}}-\ln
3)$ the singlet rung state is not
involved in the ground-state, namely $\epsilon ^{(3)}(0)\geq
0$, whereas two triplet fermi seas still have fermi boundaries at
infinity. Under such a configuration, the dressed energy potentials are
\begin{equation}
\epsilon^{(1)} (\lambda )=
-\frac{2\pi J_{\parallel}}{\sqrt{3}\gamma}\frac{ \cosh
\frac{\pi}{3}\lambda }{\cosh \pi \lambda},\, \,\,\,\epsilon^{(2)} (\lambda
)= -\frac{2\pi J_{\parallel} }{\sqrt{3}\gamma}\frac{ \sinh
\frac{\pi}{3}\lambda }{\sinh \pi \lambda}.
\end{equation}
The free energy can be given by
\begin{equation}
\frac{F(0,0)}{2L} \approx -\frac{2J_{\parallel}}{3\gamma}(\psi(1)-\psi(\frac{1}{3}))+
\frac{1}{2L}f_{{\rm b}},\label{freeE}
\end{equation}
where  
\begin{eqnarray}
f_{{\rm b}}&=&\int_{-\infty}^{\infty}\rho_{{\rm b}1}^{(1)}(\lambda)\epsilon_1^{(1)}(\lambda)\d\lambda +\int_{-\infty}^{\infty}\rho_{{\rm b}1}^{(2)}(\lambda)\epsilon_1^{(2)}(\lambda)\d\lambda.
\end{eqnarray}
The first part in (\ref{freeE}) is nothing but the standard $SU(3)$
ground state energy of the bulk. The remaining part is the boundary surface
energy for various boundary impurities.
\section{Conclusion and discussion}
\label{sec5}

In summary, we have discussed in detail the algebraic Bethe-ansatz
solution of an integrable spin ladder system based on the $SU(4)$
symmetry with boundary impurities. Five different classes of solutions
of the graded RE leading to different boundary rung interactions in
the Hamiltonian were obtained.  The Bethe-ansatz equations, the
eigenvalues of the transfer matrix and the energy spectrum were given
explicitly.  Furthermore, the three-level transfer matrices,
characterizing the different flavour sectors separately, allowed us to
embed different impurities into the system.  From the Bethe ansatz
solutions (\ref{Tbethe1})-(\ref{Tbethe3}), we  found that the
boundary impurity effects characterized by $\zeta(v_i,\xi_{\pm})$,
$\eta(\mu_j,\xi_{\pm})$ and $\Omega (w_k,\xi_{\pm})$ act indeed
non trivially on the densities of roots for the three rapidities and
thus change the ground state properties, the boundary bound states as
well as the low-lying energy spectrum. In the thermodynamic limit, the
spin gap remains almost unchanged. However, the boundary
susceptitblity and magnetization reveal novel magnetic properties for
strong and weak impurity couplings.  In strong impurity coupling, the
impurities induced by the open boundary conditions can result in
either a pure triplet or a singlet edge state due to the nature of the
pure back-scattering at the edges and the magnetic
impurities. Strictly speaking, the edge state can be a pure singlet
state in a strong antiferromagnetic boundary coupling regime whereas a
triplet state with an effective magnetic moment can exist in a strong
ferromagnetic boundary coupling regime. Correspondingly, the one point
correlation function for strong antiferromagnetic boundary impurities
tends to the singlet eigenvalue $-\frac{3}{4}$, whereas for strong
ferromagneitc impurity coupling it tends to the triplet eigenvalue
$\frac{1}{4}$. This behaviour may be observed in experiments due to
different boundary magnetic moments.  Although  the TBA solution  of the $SU(4)$ ladder model (\ref{Ham}) predicts the quantum  phase diagram in  good agreement with experimental results for the  strong coupling compounds,
the full finite temperature  thermodynamic properties of the model remain to be calculated. 

\begin{ack} A.F. and X.W.G. would like to thank J. Links, I. Roditi, R.A.~R\"{o}mer  and
Z.-J. Ying for  helpful discussions.
M.T.B., X.W.G. and H.Q.Z  thank the Australian Research Council for financial support.
 A.F. and A.P.T.  thank CNPq and  FAPERGS for financial support. 
\end{ack}


\end{document}